\documentclass[onecolumn,showpacs,nobibnotes,nofootinbib,12pt]{revtex4-1}
\usepackage{amsmath,amssymb}
\usepackage{wasysym}
\usepackage{graphicx}
\newcommand{\be}{\begin{equation}}
\newcommand{\ee}{\end{equation}}
\newcommand{\bea}{\begin{eqnarray}}
\newcommand{\eea}{\end{eqnarray}}
\newcommand{\f}{\frac}

\newcommand{\nn}{\nonumber}

\begin{document}
\author{Bin Wu}
\email{binwu@physik.uni-bielefeld.de}
\affiliation{Faculty of Physics, University of Bielefeld, D-33501 Bielefeld, Germany}

\title{On holographic thermalization and gravitational collapse of massless scalar fields}% in the Poincare patch of $AdS_5$
\begin{abstract}
In this paper we study thermalization in a strongly coupled system via AdS/CFT. Initially, the energy is injected into the system by turning on a spatially homogenous scalar source coupled to a marginal composite operator. The thermalization process is studied by numerically solving Einstein's equations coupled to a massless scalar field in the Poincare patch of $AdS_5$. We define a thermalization time $t_T$ on the AdS side, which has  an interpretation in terms of a spacelike Wilson loop $\left< W(l \simeq\f{1}{T}) \right>$ in CFT. Here  $T$ is the thermal equilibrium temperature. We study both cases with the source turned on in short($\Delta t \lesssim \f{1}{T}$) and long($\Delta t \gtrsim \f{1}{T}$) durations. In the former case,  the thermalization time  $t_T = \f{g_t}{T} \lesssim\f{1}{T}$ and the coefficient $g_t \simeq 0.73$ in the limit $\Delta t \lesssim\f{0.02}{T}$. In the latter case, we find double- and multiple-collapse solutions, which may be interpreted as the gravity duals of two- or multi-stage thermalization in CFT. In all the cases our results  indicate that such a strongly coupled system thermalizes in a typical time scale $t_T\simeq\f{O(1)}{T}$.
\end{abstract}
\maketitle
%
%%%%%%%%%%%%%%%%%%%%%%%%%%%%%%%%%%%%%%
%
%
\section{Introduction}
Knowledge about thermalization in strongly coupled systems is important for us to understand the results from relativistic heavy-ion collisions at RHIC and LHC and some non-equilibrium processes in the early universe. The AdS/CFT correspondence\cite{Maldacena,Witten} provides us with an elegant tool for studying  the strongly coupled large $N_C$  super Yang-Mills (CFT). The gravity dual of static plasma  in 4-dimensional Minkowski space($M^4$)  is  the AdS black brane(black hole) solution in the Poincare patch of $AdS_{5}$.  And studying gravitational collapse of matter fields in the bulk of $AdS_5$ helps understand the thermalization process on the CFT side. 

To study thermalization by AdS/CFT, one needs first to choose some far-from-equilibrium states. Such states can be prepared by injecting energy into the CFT vacuum. This can be done by turning on sources for some boundary operators\cite{holographicReconstruction}. Examples for such boundary sources are the boundary metric coupled to the boundary stress energy tensor\cite{CY01} or scalar sources coupled to marginal scalar composite operators\cite{Minwalla}. Another kind of initial states are obtained by reconstructing some bulk metric based on the knowledge of the CFT data\cite{holographicReconstruction}. One of this kind is the double shockwave metric\cite{JanikPeschanski2005,Yuri:2008, GrumillerPaul, Gubser:2008pc,AlvarezGaume:2008fx, Lin:2009pn, Yuri:2009,Kovchegov:2009du, CY02, withPaul,KiritsisTaliotis}, motivated by understanding thermalization in heavy-ion collisions(see \cite{CasalderreySolana:2011us}  for a recent review). Of course one can also start from the gravity side by constructing consistent initial conditions for Einstein's equations. For this kind of initial conditions, the interested reader is referred to \cite{Heller:2011} for boost-invariant cases and \cite{Bizon01,Bizon02,Garfinkle01,Garfinkle02,Gubser:2012} for the global patch of $AdS_{d+1}$. Given initial conditions and suitable boundary conditions, thermalization in strongly coupled CFT may be studied by solving Einstein's equations in the bulk of $AdS_5$.

In this paper, we study  thermalization of a spatially homogenous system in CFT\cite{Esko:1999,LinShuryak, Minwalla, holographicThermal,Galante:2012pv,Erdmenger:2012xu}. The energy is injected into the system by turning on a spatially homogenous scalar source $\phi_0(t)$ for a marginal scalar composite operator $\hat{O}$. The cases of weak fields were first tackled by the authors of Ref. \cite{Minwalla} using perturbative techniques. We will deal with more general cases using a numerical method similar to that used in the global patch of $AdS_{d+1}$\cite{Bizon01,Bizon02,Garfinkle01,Garfinkle02}. In CFT the Lagrangian of the system in general  takes the following form
\be
S = \underbrace{S_{CFT} + \int d^4x \phi_0(t) \hat{O}(t)}_{\text{thermalization}} + \underbrace{\int d^4x A_{0}^{a} J_a}_{\text{probes}},\label{equ:actionCFT}
\ee
where $S_{CFT}$ is the CFT action, $A_{0}^a$ and $J_a$ denote all the other external sources and currents(operators). On the gravity side, the bulk metric is determined by solving the gravitational equations coupled to the massless scalar field $\phi$ corresponding to $\phi_0$. And the back-reaction of all the other fields $A^a$ corresponding to the boundary sources $A_0^{a}$ is assumed to be negligible. The expectation values of $J_a$ and their correlators can be calculated by solving the equations of motion of those weak fields in the bulk metric\cite{Witten}. They are also essential for understanding the details of thermalization in a coordinate-independent way\cite{holographicThermal}. %In the following we will focus on solving the bulk metric from Einstein-Klein-Golden equations but define a thermalization time with an interpretation in terms of the spacelike Wilson loop in CFT.

In the introduction we summarize our results. The metric in Schwarzschild(Poincare) coordinates can be written in the form%\footnote{In Appendix \ref{app:B}, we also give the equations of motion in Eddington-Finkelstein coordinates.}
\be\label{equ:FeffermanGraham}
ds^2 = \f{1}{u^2}\left( -f e^{-2 \delta} dt^2 + f^{-1} du^2 + d\vec{x}^2 \right),
\ee
where $f$ and $\delta$ are functions of $t$ and $u$ only. And the AdS black hole metric is given by
\be
f_{bh} = 1-\f{u^4}{u_0^4},\text{ and } \delta = 0,\label{equ:adsstaticbh}
\ee
where $u_0=\f{1}{\pi T}$ and $T$ is the thermal equilibrium(Hawking) temperature. The energy is injected into the system according to\cite{holographicReconstruction}
\be
{{\dot{T}}^{ (4) } }_{00} = \left< \hat{O}\right> \dot{\phi_0},\label{equ:energyConservation}
\ee 
where $ \left< \hat{O}\right>$ is the expectation value of $\hat{O}$. The source $\phi_0(t)$ explicitly takes the following form
\be
\phi_0(t) = \f{\epsilon}{a} e^{-a t^2},\label{equ:phi0t}
\ee 
where $\epsilon$ and $a\equiv \f{1}{\Delta t^2}$ are two parameters and $\Delta t$ characterizes the duration of the source being turned on. However, our qualitative conclusions should hold regardless of the source's shape. The source induces an ingoing wave of the scalar field in the bulk, which will eventually collapse to form a black hole. We will not define the thermalization time $t_T$ in terms of the formation time of the apparent horizon because it is  coordinate-dependent\cite{WaldGR,Hubeny:2011}. Instead, we define $t_T$ as the moment when the AdS black hole metric is established in the most part of the bulk that is causally connected to the boundary. $t_T$ can be interpreted as the scale-dependent thermalization time\cite{holographicThermal} defined by a spacelike Wilson loop $\left< W(l)\right>$ with $l\sim \f{1}{T}$ in CFT. With such a definition, our results respect the following scaling invariance
\be
(a,\epsilon, t_T, T) \to (\lambda^2 a, \lambda^2 \epsilon, t_T/\lambda, \lambda T)
\ee
with $\lambda >0$, which gives
\be
t_T = \f{g_t}{T}.\label{equ:tTg}
\ee
$g_t$ is scale-dependent\cite{holographicThermal} and in this paper we only discuss thermalization with $l \sim \f{1}{T}$. In the following, we will show that $g_t=O(1)$ in this case.
\begin{figure}
\begin{center}
\includegraphics[width=16cm]{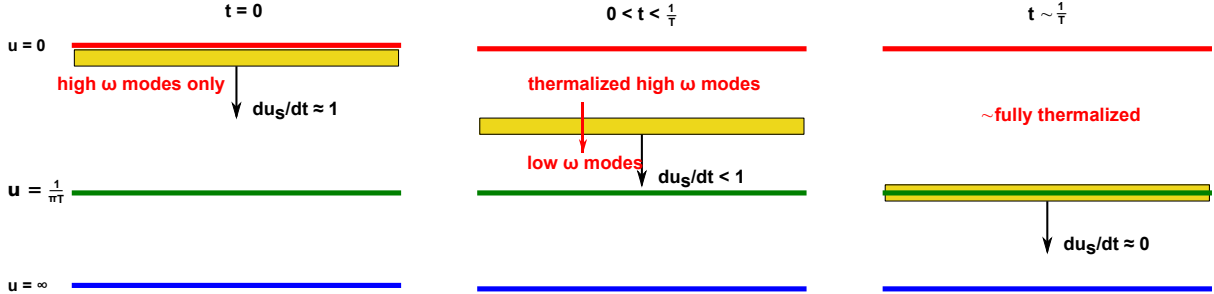}
\end{center}
\caption{Narrow wave($\Delta t \lesssim \f{1}{T}$). At $t \sim \Delta t$, the source on the boundary CFT is turned off and a narrow ingoing wave with a width $\Delta u \sim \Delta t $ is induced near the boundary $u=0$. The wave starts to propagate at the speed of light($du_s/dt\simeq1$) in the bulk and leaves behind a submanifold$(u<u_s)$ equipped with the AdS black hole metric. It propagates more slowly in the deeper interior of $AdS_5$. At $t \sim \f{1}{T}$, $du_s/dt\simeq0$ and the AdS black hole metric is established in the most part of the bulk that is causally connected to the boundary.  
}\label{fig:topdown}
\end{figure} 

In the cases with $\f{\epsilon}{a}\lesssim 1$, the boundary source induces  a narrow wave with $\Delta t \equiv \f{1}{\sqrt{a}} \lesssim \f{1}{T}$ in the bulk. Our results are illustrated in Fig. \ref{fig:topdown}. We find that $g_t\lesssim 1$ for all the narrow waves and $g_t \simeq 0.73$ if $\Delta t \lesssim \f{0.02}{T}$. In CFT, the interpretation of our results is as follows: the system, after the source is turned off, thermalizes in a time scale $t_T \sim \f{1}{T}$ in a top-down manner\cite{LinShuryak,holographicThermal}. This is in sharp contrast with bottom-up thermalization in perturbative QCD\cite{bottomup,Kurkela:2011}, in which soft gluons equilibrate more quickly than hard gluons. In CFT the interaction is equally efficient for soft and hard modes because of the vanishing beta function. At $t\sim \Delta t$, only time-like high momentum($\omega\sim \f{1}{\Delta t} = \sqrt{a}$) modes are present in the system(see \cite{energylossandpT} for a similar interpretation in the case of a classical string and \cite{HattaIancuAl} for an $R$ current). Those high $\omega$ modes split into low $\omega$ modes very rapidly\cite{HattaIancuAl}. In the meanwhile the interaction is so efficient  that the (remaining) high $\omega$ modes equilibrate in a time $\sim\f{1}{\omega}$. Such a splitting-equilibration continues  from higher $\omega$ to lower $\omega$ modes.  At $t \sim \f{1}{T}$ all the modes with $\omega \gtrsim T$ achieve thermal equilibrium. 
\begin{figure}
\begin{center}
\includegraphics[height=6cm]{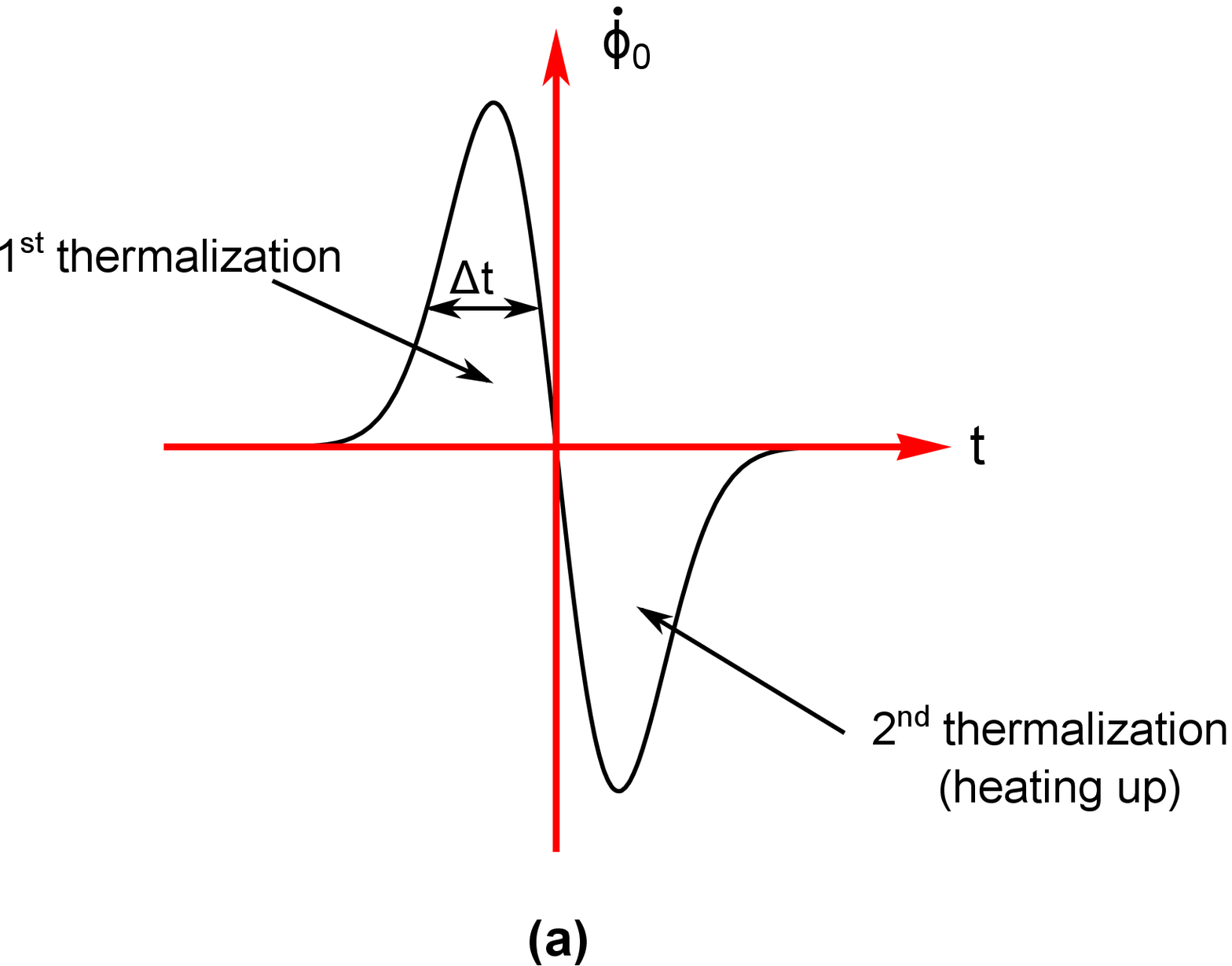}
\includegraphics[height=6cm]{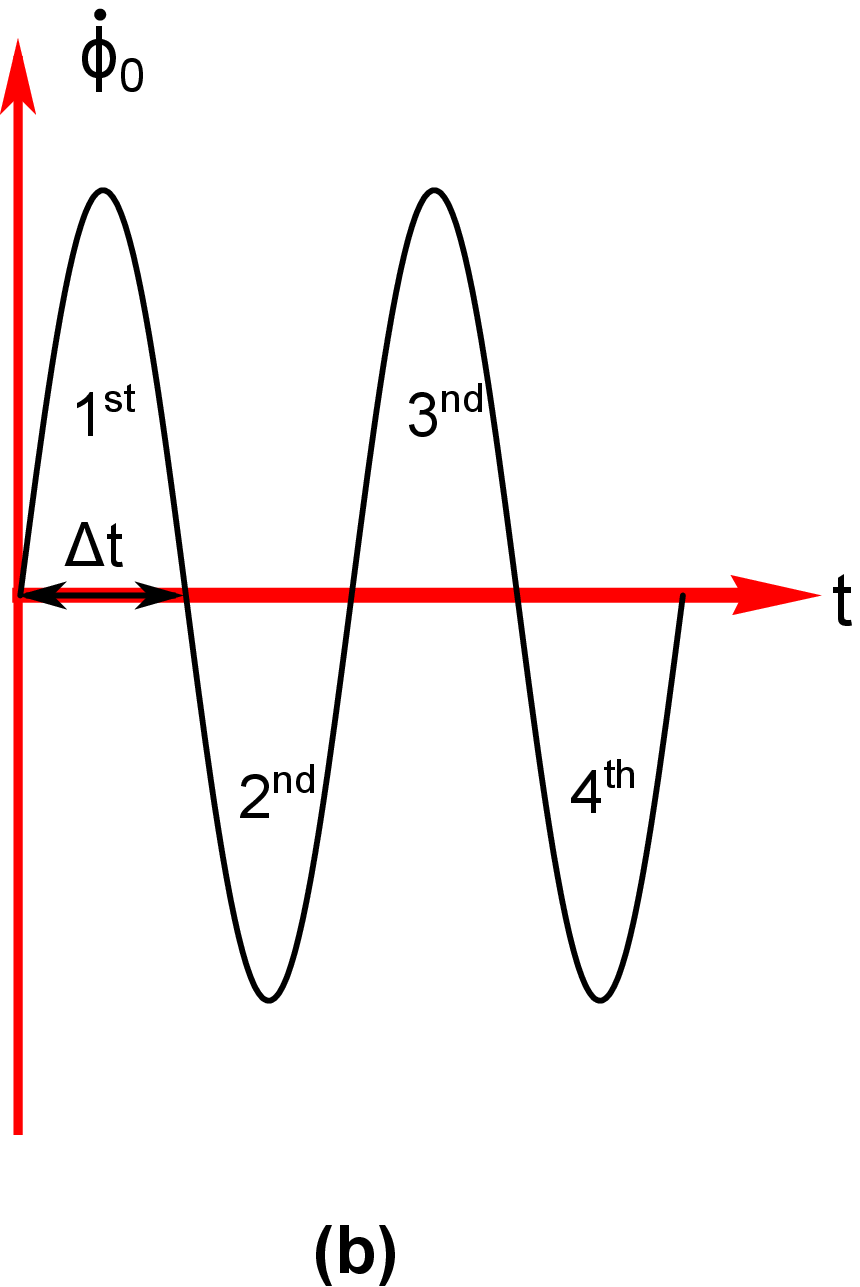}
\end{center}
\caption{Two-stage and multi-stage thermalization. Fig. (a): The energy is injected into the CFT vacuum by the two pulses of $\dot{\phi}$. If  $\Delta t \gtrsim \f{1}{T_L}$, the system first thermalizes at a lower temperature $T_L$ at $t \simeq 0$. Then, the second pulse heats up the plasma to a higher temperature $T_H$. Fig. (b) shows a source for a possible multi-stage thermalization.
}\label{fig:2stagesThermalization}
\end{figure}

%%
%%
%\begin{figure}
%\begin{center}
%\includegraphics[width=8cm]{e03b01E}
%\includegraphics[width=8cm]{e03b01f}
%\end{center}
%\caption{Broad wave($\Delta t \gtrsim \f{1}{T}$): disturbed thermalization. For this example, $\Delta t = 3.16$ and $\f{1}{T} = 2.29$. Fig. (a): the energy density of the scalar field starts to increase from about $t \sim -2 \Delta t$ to $t = 0$. However, the peak of the energy density can not go much further inside the interior of $AdS_5$ and rapidly aggregate around $u\sim \f{1}{\pi T}$ at $t\simeq0$ when the source is turned off on the boundary. Fig. (b) shows f at different times. The red solid curves staring from top right are the results from $t = -5$ to $-1$. f for $u< $ at $t\simeq 0$ is that of the AdS black hole metric.
%}\label{fig:e03b01}
%\end{figure} 
%%

In the cases with $\f{\epsilon}{a}\gtrsim 1$, the induced waves are broad and we do not evaluate the exact value of $g_t$ in (\ref{equ:tTg}). The typical thermalization time $t_T$ shows up in a way illustrated in Fig. \ref{fig:2stagesThermalization}. The system is expected to thermalize in a time $t_T\sim \f{1}{T}$. Therefore, if the source is turned on and off several times with period $\Delta t \sim \f{1}{T}$, the system should achieve thermal equilibrium within each time interval $\Delta t$. We refer to such a thermalization pattern as a two-stage(Fig. \ref{fig:2stagesThermalization}(a)) or multiple-stage(Fig. \ref{fig:2stagesThermalization}(b)) thermalization. This is the interpretation of the double- or multiple-collapse solutions that we find in the bulk. According to eq. (\ref{equ:energyConservation}), the energy is injected into the system by the two pulses of $\dot{\phi}$ as illustrated in Fig. \ref{fig:2stagesThermalization}(a).  After the source is turned off at $t \gtrsim 2 \Delta t$, the scalar field will eventually collapse. If  $\f{\epsilon}{a} \gtrsim 3.0$, at $t\simeq0$ the falling of the scalar field results in a submanifold defined by $u\leq 0.98 u_0$, which is equipped with the AdS black hole metric with the Hawking temperature $T=T_L$. We interpret it as an intermediate thermal equilibrium state with $T=T_L$ in CFT.  And it is called the first collapse even though  the trapped region has not formed yet(in Schwarzschild coordinates) at this time.  Then the final collapse at $t\sim 2 \Delta t$ is naturally interpreted as the second stage of thermalization(heating-up process) in CFT. The criteria for obtaining such double-collapse solutions is $\Delta t \gtrsim \f{1}{T_L}$. If $\Delta t > \f{1}{T_L}$, the system thermalizes in a time $\sim \Delta t > \f{1}{T_L}$ simply because the thermal equilibrium is destroyed by the continuous injection of high $\omega$ modes from top down before $t=0$. A multiple-collapse solution can be defined in a similar way and we find that the above criteria is also parametrically true for obtaining multiple-collapse solutions induced by a periodic source(see Fig. \ref{fig:2stagesThermalization}(b)). Therefore, we conclude that in such a strongly coupled system the typical thermalization time is $t_T \sim \f{1}{T}$.

This paper is organized as follows. In Sec. \ref{sec:eom}, we derive the equations of motion for a massless scalar field coupled to gravity in the Poincare patch of $AdS_5$. The numerical schemes in addition to initial conditions and boundary conditions are discussed in Sec. \ref{sec:numerics}.  In this section, we also define the thermalization time $t_T$. 
Our numerical results are presented in Sec. \ref{sec:results}. In Sec. \ref{sec:discussion}, we briefly conclude. In Appendix \ref{app:EddingtonFinkelstein}, we give the equations of motion in Eddington-Finkelstein coordinates. The details of our numerical methods are presented in Appendix \ref{app:numerics}.

%
%%%%%%%%%%%%%%%%%%%%%%%%%%%%%%%%%%%%%%
%
%
 \section{ Einstein-Klein-Gordon equations}
\label{sec:eom}
On the AdS side, we need to calculate the back-reaction of a massless scalar field to the bulk geometry. The bulk action  in $AdS_{d+1}$ corresponding to the first two terms on the right-hand side of (\ref{equ:actionCFT}) is given by
\begin{equation}
S=\f{1}{2\kappa_{d+1}^2}\left\{ \int
d^{d+1}x\sqrt{-g}\left\{ R-2\Lambda - 2 \left(\partial \phi \right)^2  \right\} + 2 \int_{\partial M} d^dx \sqrt{\gamma} K \right\},\label{equ:action}
\end{equation}
where $\Lambda=-{d(d-1)\over 2L^2}$ for $AdS_{d+1}$, $\kappa_5^2 = \f{4 \pi^2 L^3}{N_c^2}$, $L$ is a parameter of dimension of length, $\gamma$ is the induced metric on the boundary and $K$ is the trace of the extrinsic curvature of the boundary. %% The action is invariant under the dilation 
%\be
%x^a \rightarrow \lambda x^a,~~\phi\rightarrow \phi(x^a)\label{equ:dilation}
%\ee
%with $\lambda$ a constant.
We need to solve the following Einstein-Klein-Gordon equations
\bea
&&\partial_a\left(\sqrt{-g} g^{ab} \partial_b \phi \right)=0,\\
&&R_{ab}-\f{1}{2} g_{ab} R - \f{d(d-1)}{2L^2} g_{ab} = T_{ab},
\eea
where $T_{ab}$ is the stress tensor of the scalar field, which is given by
\be
T_{ab}= 2 \partial_a \phi \partial_b \phi - g_{ab}  \left(\partial \phi \right)^2.
\ee
In the following, we take $L=1$ and $d=4$.
%In the onshell action of the $AdS_{d+1}$, $ R-2\Lambda - 2 \left(\partial \phi \right)^2=-2d+\f{4}{d-1} m^2 \phi^2$. In this paper, we only deal with massless scalar fields.

In this paper, the scalar sources are assumed to be spatially homogeneous on the boundary $M^4$. In Schwarzschild coordinates, one needs to solve the following equations of motion
\begin{subequations}\label{equ:eom}
\bea
&&\dot{V} = u^3 \left(\f{f e^{-\delta} P}{u^3}\right)^\prime,\label{equ:Vdot}\\
&&\dot{P} = \left( f e^{-\delta} V \right)^\prime,\label{equ:Pdot}\\
&&\dot{f} = \f{4}{3} u f^2 e^{-\delta} V P,\label{equ:fdot}\\
&&\delta^\prime=\f{2}{3} u \left(  V^2 + P^2 \right),\label{equ:deltap}\\
&&f' = \f{2}{3} u f \left(V^2+ P^2\right) +\f{4}{u}\left( f-1\right),\label{equ:fp}
\eea
\end{subequations}
where the derivatives with respect to $t$ and $u$ are denoted respectively by overdots and primes, $P\equiv \phi^\prime$, $V\equiv f^{-1} e^\delta \dot{\phi}$ and we take (\ref{equ:fp}) as a constraint equation. If $V$ and $P$ vanish in a submanifold near the boundary of $AdS_5$, the only solution to (\ref{equ:deltap}) and (\ref{equ:fp}) is
\be
f = 1 - \f{u^4}{u_0^4},~~\delta = 0,~~V=0~~\mbox{and}~~P=0,\label{equ:adsbhasy}
\ee
which is Birkhoff's theorem\cite{WaldGR} in such a spatially homogeneous case in $AdS_5$. The equations of motion above are invariant under the following scaling transformation
\bea\label{equ:dilation}
&&\phi(x^a) \rightarrow \tilde{\phi}=\phi(\lambda^{-1} \tilde{x}^a), ~~V\rightarrow \lambda \tilde{V},~~P\rightarrow \lambda \tilde{P},\nn\\\
&&x^a \rightarrow \tilde{x}^a=\lambda x^a,~~f(x^a)\rightarrow \tilde{f}(\tilde{x}^a)=f(\lambda^{-1}\tilde{x}^a),~~\delta(x^a)\rightarrow \tilde{\delta}(\tilde{x}^a)=\delta(\lambda^{-1}\tilde{x}^a).\eea

%
%%%%%%%%%%%%%%%%%%%%%%%%%%%%%%%%%%%%%%
%
\section{Gravitational collapse of massless scalar fields}\label{sec:numerics}
In this paper, we study the response of the vacuum/plasma to an external scalar source in CFT. On the gravity side, the scalar source provides the boundary conditions and  the vacuum AdS/the AdS black hole metric provides the initial conditions for solving  eq. (\ref{equ:eom}). We aim to see how a submanifold near the boundary of $AdS_5$ equipped with the AdS black hole metric in (\ref{equ:adsstaticbh}) forms by  gravitational collapse of a massless scalar field. 
 \subsection{Initial conditions}
 Before the scalar source is turned on, the scalar field is assumed to vanish in the bulk. In this case, by Birkhoff's theorem one has
\be
f_{bh}=1-\f{u^4}{u_0^4},~~\delta=0,~~V=0~~\mbox{and}~~P=0,\label{equ:adsbh}
\ee
which is the AdS black hole metric in eq. (\ref{equ:adsstaticbh}), the gravity dual of static plasma. By holographic renormalization, the stress tensor of the boundary CFT is given by\cite{Kraus}
\be
{T^{(4)\mu}}_{\nu} = \f{3}{8 }  N_c^2 \pi^2 T^4 \mbox{diag} \{-1,1/3,1/3,1/3 \}.
\ee
Taking $T\to 0$, one gets the vacuum AdS metric, dual to the CFT vacuum, as follows
 \bea
f_{vac}=1,~~\delta=0,~~V=0~~\mbox{and}~~P=0.\label{equ:initial}
\eea
Both (\ref{equ:initial}) and (\ref{equ:adsbh}) will be used as initial conditions for solving (\ref{equ:eom}) in this paper.
\subsection{Boundary conditions}
\begin{figure}
\begin{center}
\includegraphics[width=8cm]{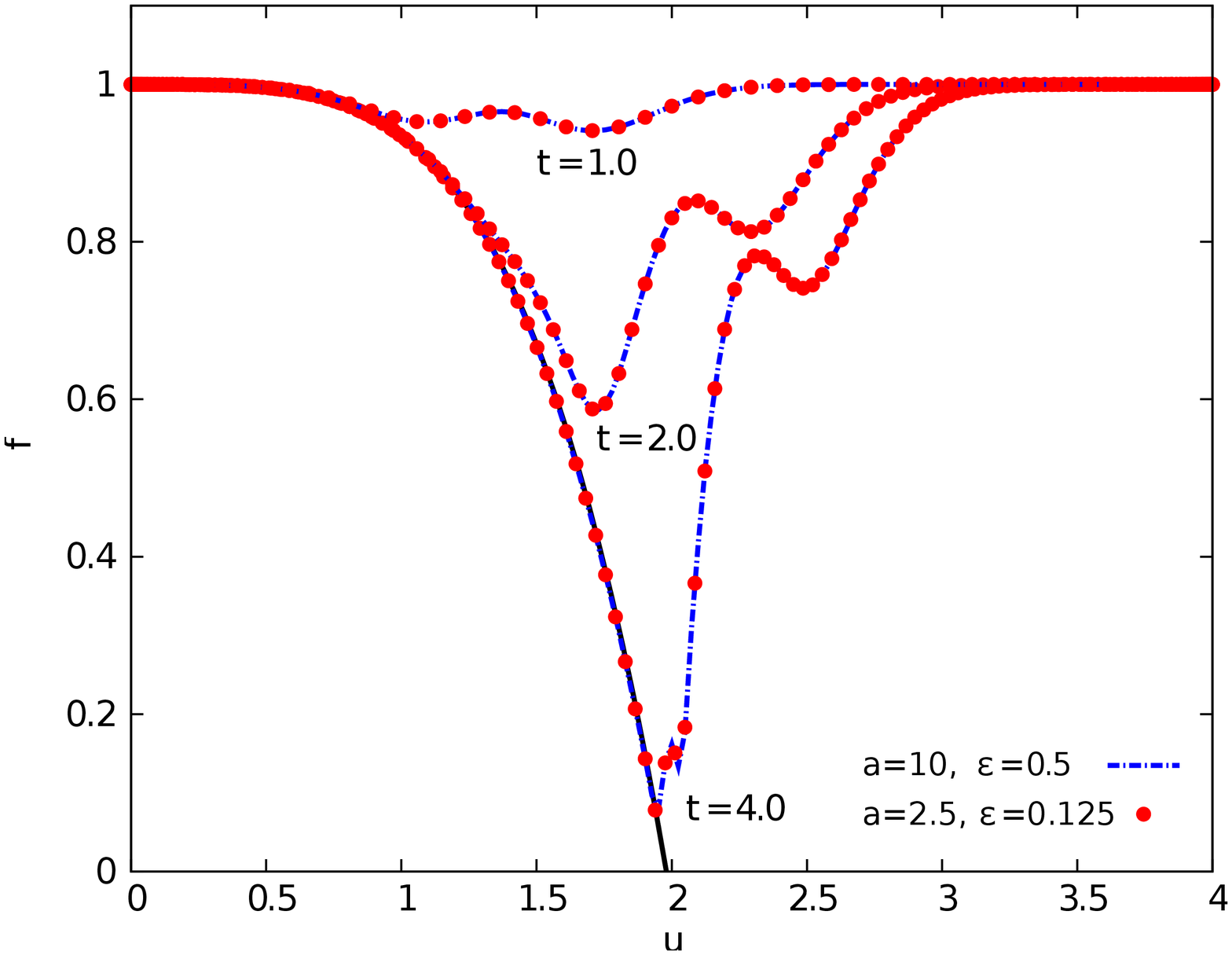}
\includegraphics[width=7.8cm]{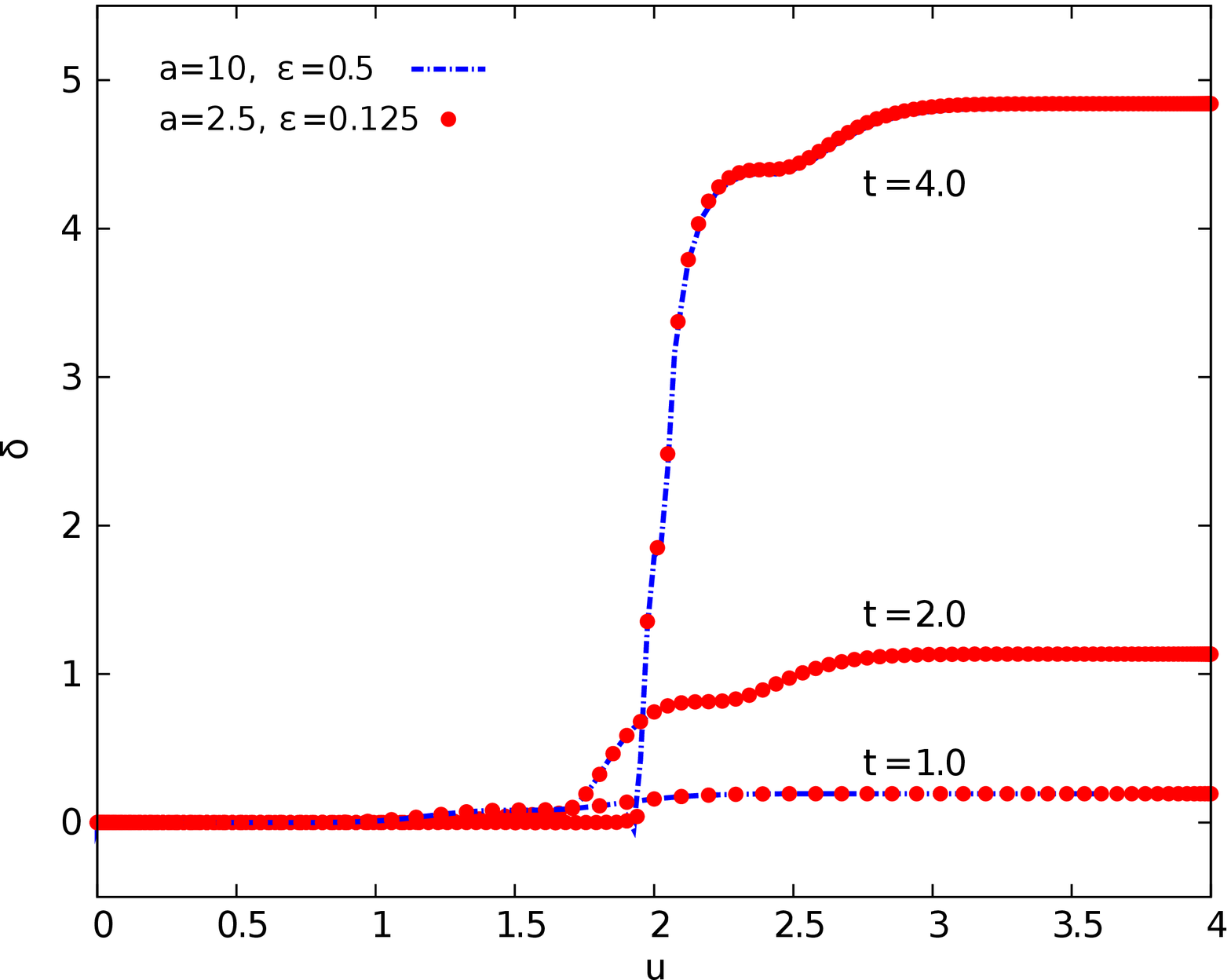}
\end{center}
\caption{Usage of the scaling transformation in (\ref{equ:dilation}). Here, we show two numerical solutions with boundary conditions respectively given by $(a, \epsilon)=(2.5, 0.125)$ and $(a, \epsilon) =  (10, 0.5)$. In these two figures the dashed curves show the metric functions $f$ and $\delta$ of the solution with $(a, \epsilon)=(10, 0.5)$ under the scaling transformation($\lambda = 2$). They are the same as those with $(a, \epsilon)=(2.5, 0.125)$.  This example illustrates that for the equivalent solutions in (\ref{equ:dilationEqual}) we need only to calculate one of them and get the rest by the scaling transformation in (\ref{equ:dilation}).
}\label{fig:dilationPoincare}
\end{figure}

The boundary condition for solving (\ref{equ:deltap}) is given by
\be
 \delta(t,0)=0.\label{equ:deltabc}
\ee
The boundary conditions for solving (\ref{equ:Vdot}) and (\ref{equ:Pdot}) are given by the scalar source in eq. (\ref{equ:phi0t}) in the boundary CFT, which is rewritten in the following form 
\be
V(t,0)=-2 t \epsilon e^{-a t^2}\text{ and }P(t,\infty) = 0.\label{equ:phi0}
\ee
The above boundary conditions give  a unique solution in the bulk. Another solution can be obtained by replacing $(a, \epsilon)$ with  $ (\lambda^2 a, \lambda^2 \epsilon)$  in (\ref{equ:phi0}). These two solutions are related to each other  by the scaling transformation in eq. (\ref{equ:dilation}).  In the following, we denote such an equivalence briefly by
\be
(a, \epsilon) \cong (\lambda^2 a, \lambda^2 \epsilon).\label{equ:dilationEqual}
\ee
As a result, we only need to study the dependence of solutions either on $a$ or $\epsilon$(see Fig. \ref{fig:dilationPoincare} for an example). 

\subsection{Numerical scheme}
The equations of motion in eq. (\ref{equ:eom}) can be solved numerically either by the Chebyshev pseudo-spectral method\cite{withPaul} or the finite difference method\cite{MortonMayers}(see Appendix \ref{app:numerics} for a detailed description of our numerical methods). Given $f$, $\delta$, $V$ and $P$ at $t = t_n$, we first calculate $V$, $P$ and $f$ at the next time step $t_{n+1}$ by solving (\ref{equ:Vdot}), (\ref{equ:Pdot}) and (\ref{equ:fdot}). We use the third order Adams-Bashforth method as our time-marching scheme. Then, $\delta$  at $t_{n+1}$ is obtained by solving eq. (\ref{equ:deltap}). Given the initial conditions in (\ref{equ:initial})/(\ref{equ:adsbh}), the bulk metric at late times can be calculated by repeating the above two steps.

In numerical simulations, one needs only to study the evolution of the scalar field in the bulk region $0\leq u\leq u_{max}$ with $u_{max}$ being some bulk cutoff.  $V = 0$ and $P = 0$ in the bulk region in which the scalar field has not reached yet. From (\ref{equ:eom}), it is easy to show that $f$ is independent of $t$ and $\delta$ is independent of $u$ in this region. The geometry in this region can not influence the propagation of the scalar in the bulk. Therefore, one can arbitrarily choose $u_{max}$ to be any point in this region. In this case, the boundary condition at $u=\infty$ in (\ref{equ:phi0}) is replaced by $P(t, u_{max}) = 0$. On the other hand, it helps to save computation time by choosing $u_{max}$ close to the deepest region that the scalar field can reach within the thermalization time $t_T$. Using trial and error, we find that it is sufficient to choose $u_{max} \simeq 2 u_0$ or $u_{max}\simeq 3 \Delta t$ respectively for narrow waves or broad waves discussed in the next section.

%
%\subsection{Locating apparent horizons}
%%
%Let us choose the future-pointing time-like unit normal vector $n^a$ and the outward-pointing unit normal vector $s^a$ to be
%\bea
%&&n^a = u f^{-1/2} e^{\delta}\delta^a_0,\\
%&&s^a =- u f^{1/2}\delta^a_u.
%\eea
%The outgoing and ingoing null vectors are given by
%\bea
%&&k^a = \f{1}{\sqrt{2}}\left( n^a + s^a\right),\\
%&&l^a = \f{1}{\sqrt{2}}\left( n^a - s^a\right).
%\eea
%The induced metric takes the following form
%\be
%m_{ab} = g_{ab}+n_a n_b-s_a s_b.
%\ee
%The expansion of the outgoing null geodesics is given by
%\be
%\Theta = \mathcal{L}_k \log m^{1/2}.
%\ee
%In our case, $m^{1/2} = u^{-3}$,
%\be
%\Theta = \f{3}{\sqrt{2}} f^{1/2}.
%\ee
%Therefore, the apparent horizon in our case is given by
%\be
%f = 0.
%\ee
%
\subsection{Thermalization time}
In this paper, we define a thermalization time $t_T$ by the first time when\footnote{As discussed below, $t_T$ is coordinate-independent but this simple form in the definition of $t_T$ is coordinate-dependent. See Appendix \ref{app:EddingtonFinkelstein} for the discussion in Eddington-Finkelstein coordinates.}
\be
f(t_T, u)\geq f_{min}~~\mbox{and}~~f(t_T,u_{min}) = f_{min},\label{equ:tT}
\ee
where $f_{min}$ is chosen to be 0.01 for the narrow waves. A similar definition of the black hole formation time is also used in the global patch by the authors of Ref.s \cite{Garfinkle01,Garfinkle02}. We also require that $f(t_T, u)$ at $u< u_{min}$ is that of the AdS black hole metric in (\ref{equ:adsbh}). The goodness-of-fit is evaluated by
\be
\sigma = \f{\sqrt{\sum\limits_{i=1}^n\left(f_{bh}(u_i)/f_i - 1\right)^2}}{n},\label{equ:sigma}
\ee
where $f_i$ is our numerical result at $u_i$ and $n$ is the number of the grid points $u_i < u_{min}$. In the following we shall show that  the energy  density of the scalar field narrowly peaks around $u=u_{min}$ when $u_{min} \gtrsim 0.98 u_0$. Before encountering the coordinate singularity at the apparent horizon(given by $f=0$), we will refer to such a state as a collapse.

\begin{figure}
\begin{center}
\includegraphics[width=8cm]{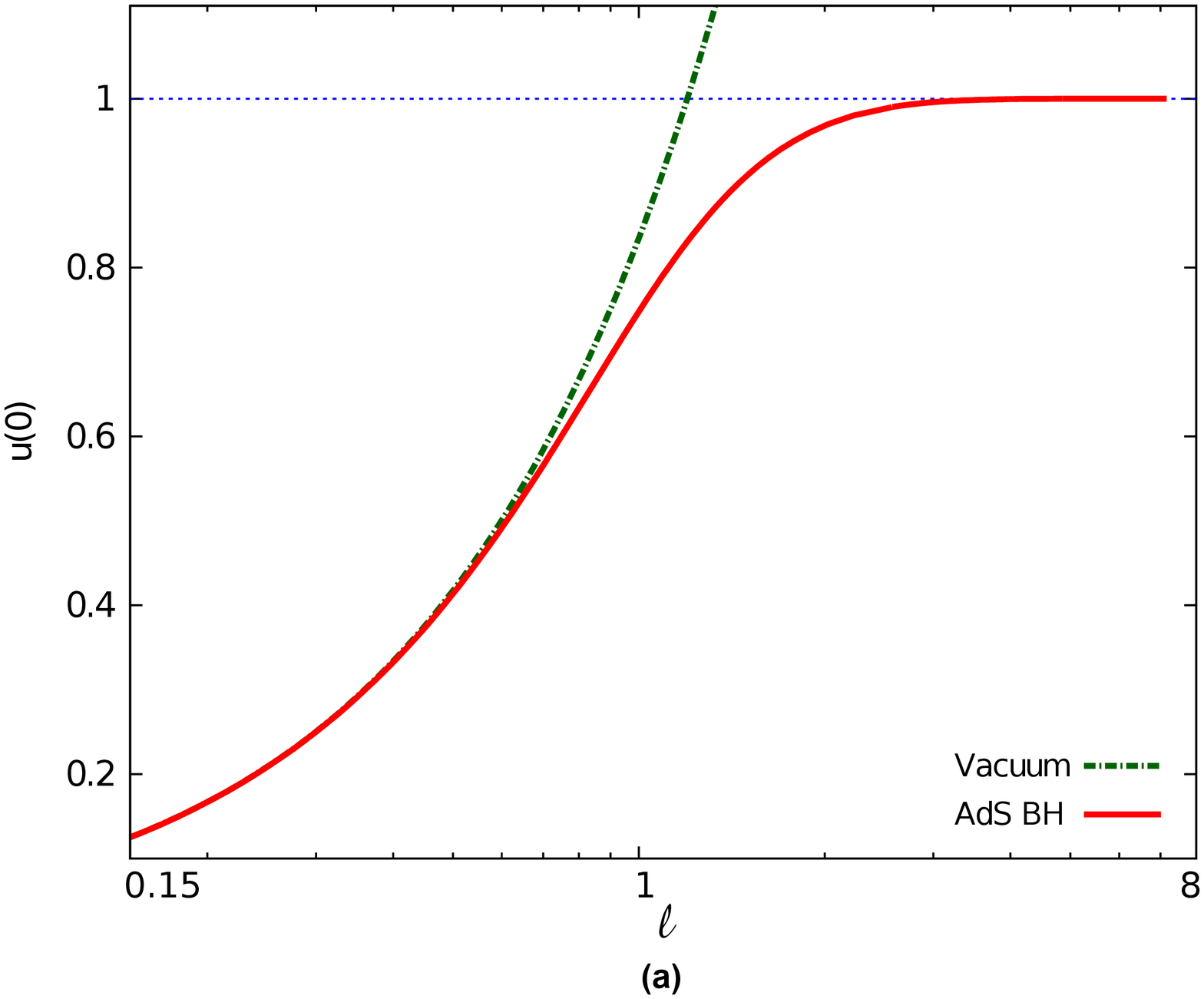}
\includegraphics[width=8cm]{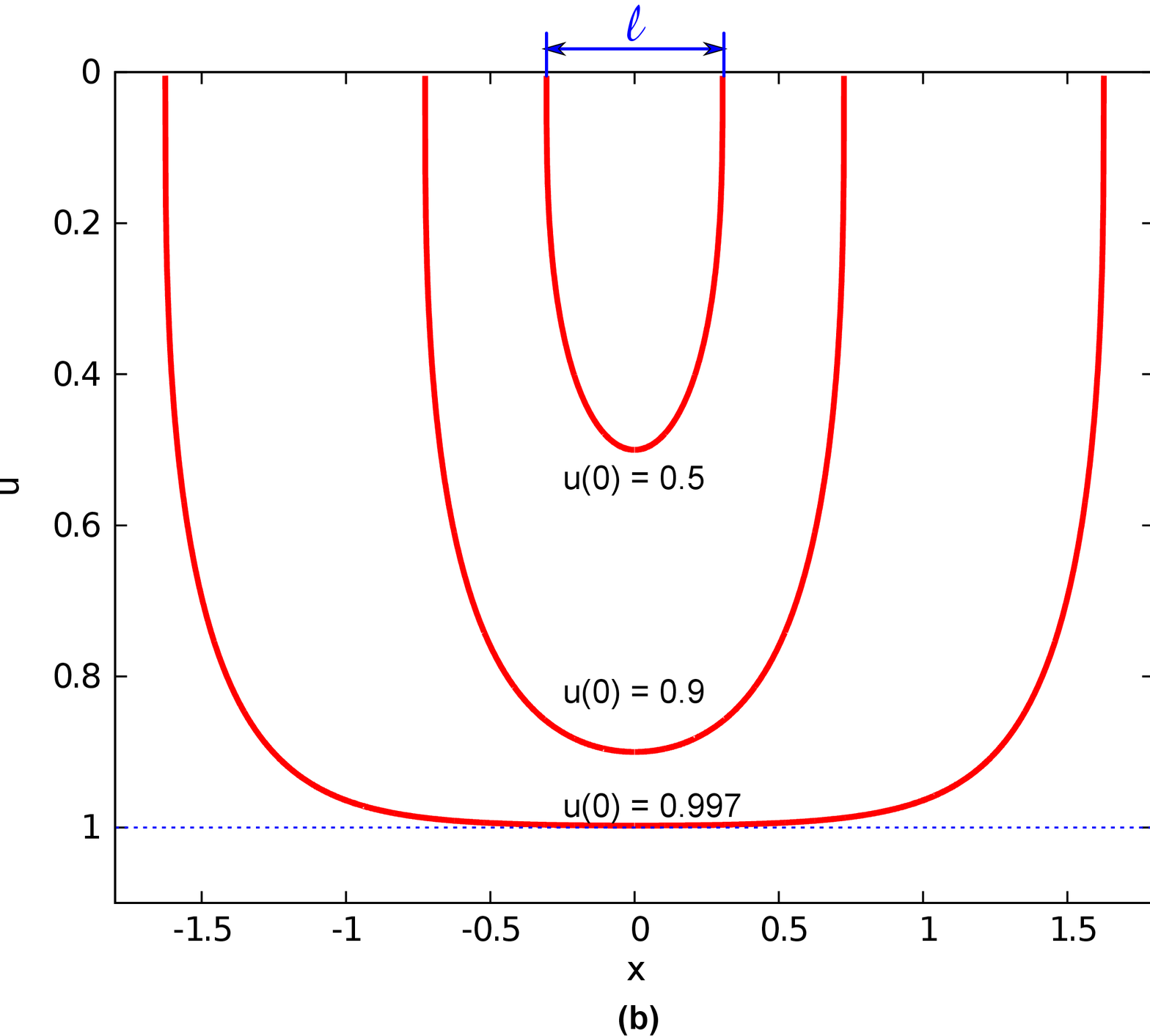}
\end{center}
\caption{Submanifold probed by a spacelike Wilson loop. Fig. (a) shows u(0) as a function of $l$ in the vacuum AdS(Vacuum) and the AdS black hole metric(AdS BH). Because of the scaling invariance in the vacuum AdS, $u(0)$ is a linear function of $l$, i.e.,  $u(0)\simeq0.835 l$(note that a logarithmic $l$-axis is used in Fig.  (a)). Fig. (b) shows the spacelike geodesics $u$ as a function of $x$ with boundary conditions $u(0) = 0.5, 0.9$ and $0.997$. Here, we take $u_0 = \f{1}{\pi T}=1$.
}\label{fig:wilsonloop}
\end{figure} 

The thermalization time $t_T$ has a coordinate-independent interpretation on the CFT side. It is the thermalization time defined by a non-local operator $\left< O(l)\right>$ at scale $l \sim \f{1}{T}$. And the $f_{min}$ dependence of $t_T$ corresponds to the scale dependence of the thermalization time discussed in Ref. {\cite{holographicThermal}}. Here we only consider the rectangular spacelike Wilson loop $\left<W(l)\right>$\cite{Maldacena:wilsonloop, holographicThermal}(see Ref. \cite{holographicThermal} for the discussion of other non-local operators).  $\left<W(l)\right>$ is a functional of spacelike geodesics which satisfy the following equations
\bea
&&d_x \left( \f{f e^{-2 \delta} d_x t}{G u^2 } \right)+\f{1}{2 G u^2 } \left( 2 \dot{\delta} f e^{-2 \delta} d_x t^2 - \dot{f} e^{-2\delta} d_x t^2 - f^{-2} \dot{f} d_x u^2    \right)=0,\\
&&d_x \left( \f{f^{-1} d_x u}{G u^2} \right)+\f{2G}{u^3} - \f{1}{2 G u^2}\left( 2 \delta' f e^{-2\delta} d_x t^2 - f' e^{-2\delta} d_x t^2 - f^{-2} f' d_x u^2 \right) = 0,
\eea
where $G\equiv  \sqrt{ 1 - f e^{-2\delta} d_x t^2 + f^{-1} d_x u^2} $, $d_x \equiv d/dx$ and the boundary conditions are given by $u(-\f{l}{2}) = 0 = u(\f{l}{2})$ and $t(-\f{l}{2})=t=t(\f{l}{2})$. As we shall see in the next section, at $t\sim t_T$ there is a sharp transition between the submanifold with vanishing $V$ and $P$ near the boundary and the rest of the bulk. In this submanifold, $\delta'$, $\dot{\delta}$ and $\dot{f}$ vanish. Therefore, it is a good approximation for us to solve the geodesic equation in the AdS black hole metric instead, which takes the following form
\be
\f{d^2}{dx^2}u + \frac{2 \left(1-2 u^4+u^8+ d_x u^2\right)}{u \left(1-u^4\right)}=0,\label{equ:u}
\ee
where $t(x) = t$ and we have taken $u_0=1$. Eq. (\ref{equ:u}) can be easily solved numerically by the shooting method starting from $u(x=0)=u(0)$ and $u'(0)=0$. As showed in Fig. \ref{fig:wilsonloop}(a), $u(0)$ is a monotonically increasing function of $l$. As a result, to probe high momentum modes($\omega\sim \f{1}{l} \gg T$)  one only needs to know the bulk metric at $u\lesssim l$ while for low momentum modes($\omega\sim \f{1}{l} \sim T$) all the information of the bulk metric at $u\lesssim u_{min}\sim u_0$ is needed. In our definition of $t_T$, $u_{min} = 0.997 u_0$. Solving eq. (\ref{equ:u}) with $u(0) = u_{min}$ gives $l = 3.16 u_0\simeq\f{1.0}{T}$. Therefore, at $t\simeq t_T$ the expectation values of $\left<W(l)\right>$ with $l\lesssim \f{1}{T}$ all reduce to thermal results(see Fig. \ref{fig:wilsonloop}(b)). In this paper, we only talk about thermalization in this sense and focus on the gravity side. A quantitative analysis like that in Ref. \cite{holographicThermal} on the CFT side is also essential for obtaining more detailed information but we leave it to future studies.

%
%%%%%%%%%%%%%%%%%%%%%%%%%%%%%%%%%%%%%%
%
\section{Results}\label{sec:results}
In this section, we discuss results with boundary sources of different values of $(a, \epsilon)$  in (\ref{equ:phi0}). According to (\ref{equ:dilationEqual}), we need only to study those parameter sets with different ratios of $\f{\epsilon}{a}$. In the following, we first fix either $\epsilon$ or $a$ for convenience of our numerical calculation. Then, general results are obtained by the scaling transformation in (\ref{equ:dilation}). We study both narrow ($\f{\epsilon}{a}\lesssim1$) and broad($\f{\epsilon}{a}\gtrsim1$) waves.
\subsection{Narrow waves$\left(\Delta t = \f{1}{\sqrt{a}} \lesssim \f{1}{T}\right)$: $\f{\epsilon}{a}\lesssim1$}
%
%In this case, results can be easily understood by the symmetries of the action in eq. (\ref{equ:action}). Besides the scaling invariance in (\ref{equ:dilation}), the action is also invariant under the following transformation
%\be
%\phi \to \phi + \Phi_0\label{equ:symmetry}
%\ee
%with $\Phi_0$ a constant. In our case, the energy is injected into the system mainly during the time interval $|t| \lesssim \Delta t = \f{1}{\sqrt{a}}$ and, therefore, on the boundary
%\be
%\phi(t,0)=\f{\epsilon}{a} e^{-a t^2} = \f{\epsilon}{a} + \epsilon t^2 + \epsilon \f{1}{2!} a t^2 + \cdot\cdot\cdot = \f{\epsilon}{a} + \epsilon t^2 + \mathcal{O}\left( \f{\epsilon}{a} \right).
%\ee
%Using (\ref{equ:symmetry}) and (\ref{equ:dilation}), one can expect that results with fixed $\epsilon$ are independent of $a$ as long as $\f{\epsilon}{a}\ll 1$. In this limit, perturbative calculations can be done in term of the expansion in $\f{\epsilon}{a}$\cite{Minwalla} and we will focus on  parameters with $\f{\epsilon}{a}\lesssim 1$.
%
%%Obviously, this argument is generally true for all the smooth boundary condition respecting the symmetry under $t\to -t$ as long as $\epsilon\Delta t^2$ is small enough. 
%Those symmetries also implies the thermalization time should be proportional to $\f{1}{T}$. %Fig. \ref{fig:tPoincare} shows that the late time location of the scalar shell(turning point of $f$) is approximately independent of $a$ if $a\geq 100$.

%
\begin{figure}
\begin{center}
\includegraphics[width=8cm]{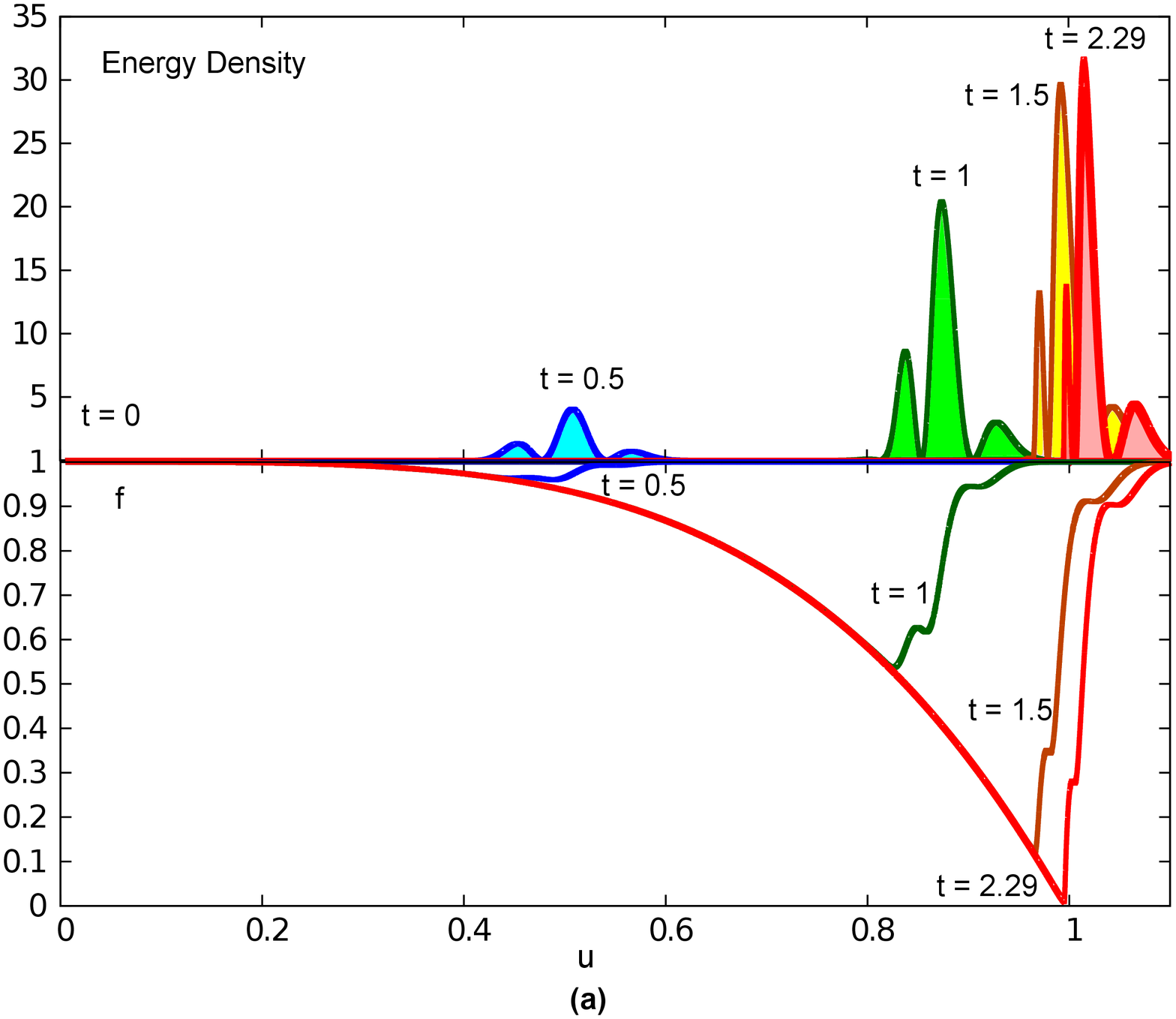}
\includegraphics[width=8cm]{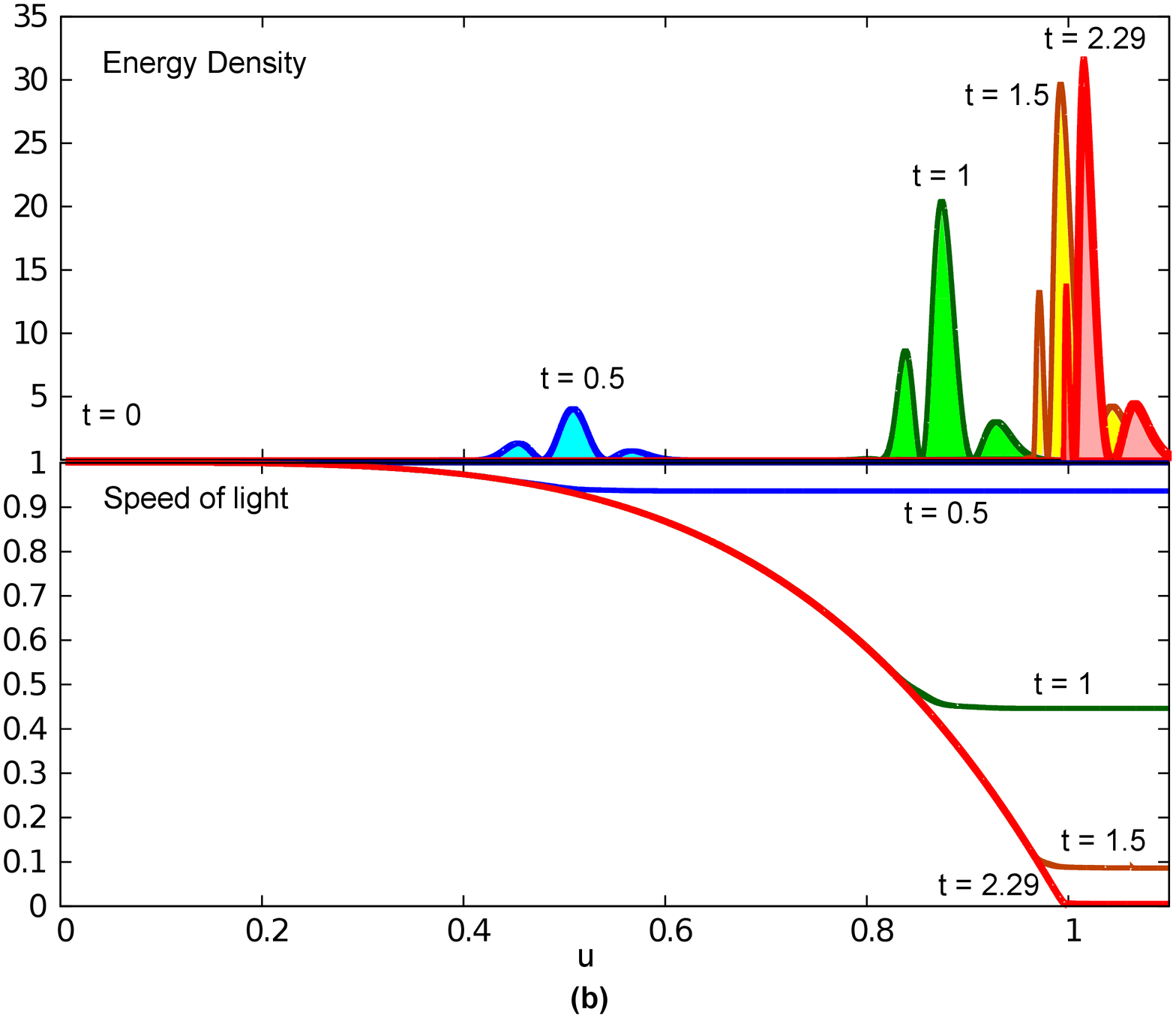}
\end{center}
\caption{An ingoing narrow wave. This is a detailed example of the falling shell illustrated in Fig. \ref{fig:topdown}. Fig. (a):  $f^2(V^2+P^2) $ and the metric function $f$ at different times. Fig. (b):  $f^2(V^2+P^2) $ and $\f{du}{dt} = f e^{-\delta}$, the speed of light-like geodesics, at different times. Here, $(a, \epsilon) = (400, 0.5)$.
}\label{fig:e001a10}
\end{figure} 
In the limit $\f{\epsilon}{a}\ll1$, the source induces a narrowly peaked ingoing wave propagating in the bulk of $AdS_5$, which has been illustrated in Fig. \ref{fig:topdown}. Let us take for example a narrow wave with $\Delta t = 0.05$ and $u_0 = 1.00$. Fig. \ref{fig:e001a10} shows the energy density of the scalar field\footnote{The energy density of the scalar field for an observer with $n^a=\f{1}{\sqrt{-g_{00}} } \delta^a_0$ is $\rho = n^a n^b T_{ab} = f \left( V^2 + P^2 \right)$. In this paper, we use $f^2 \left( V^2 + P^2 \right)$ instead in order to show the energy density at different times in the same figure.}, $f$ and the speed of light-like geodesics of the solution. At $t \simeq 2 \Delta t$, the source on the boundary CFT is turned off and a narrow ingoing wave is induced near the boundary. The wave starts to propagate with the group velocity $du_s/dt\simeq1$ in the bulk. Its group velocity becomes slower in the deeper interior of $AdS_5$. At $t = 2.29\simeq \f{0.73}{T}$ the AdS black hole metric is established in the submanifold defined by $u<0.997u_0$ and the speed of lightlike geodesics $\f{du}{dt} = f e^{-\delta} = 0.0063$ at $u\gtrsim u_0$. This is interpreted as the gravity dual of the thermalized system defined by the spacelike Wilson loop $\left< W(l\simeq \f{1}{T})\right>$ in CFT. At each time $t\gtrsim 2 \Delta t$, the bulk metric is  composed of three parts: (I) the AdS black hole metric behind the wave, (II) the Vacuum AdS metric with a time dilation, i.e., $\delta> 0$, before the wave and (III) the transition over a region $\Delta u \lesssim 4 \Delta t$ between (I) and (II) across the wave. Therefore, in this case the system in the boundary CFT thermalizes in a top-down manner\cite{holographicThermal}. Fig.  \ref{fig:e001a10}(b) shows that the speed of light becomes slower at later times in region (II).
 
\begin{figure}
\begin{center}
\includegraphics[width=8cm]{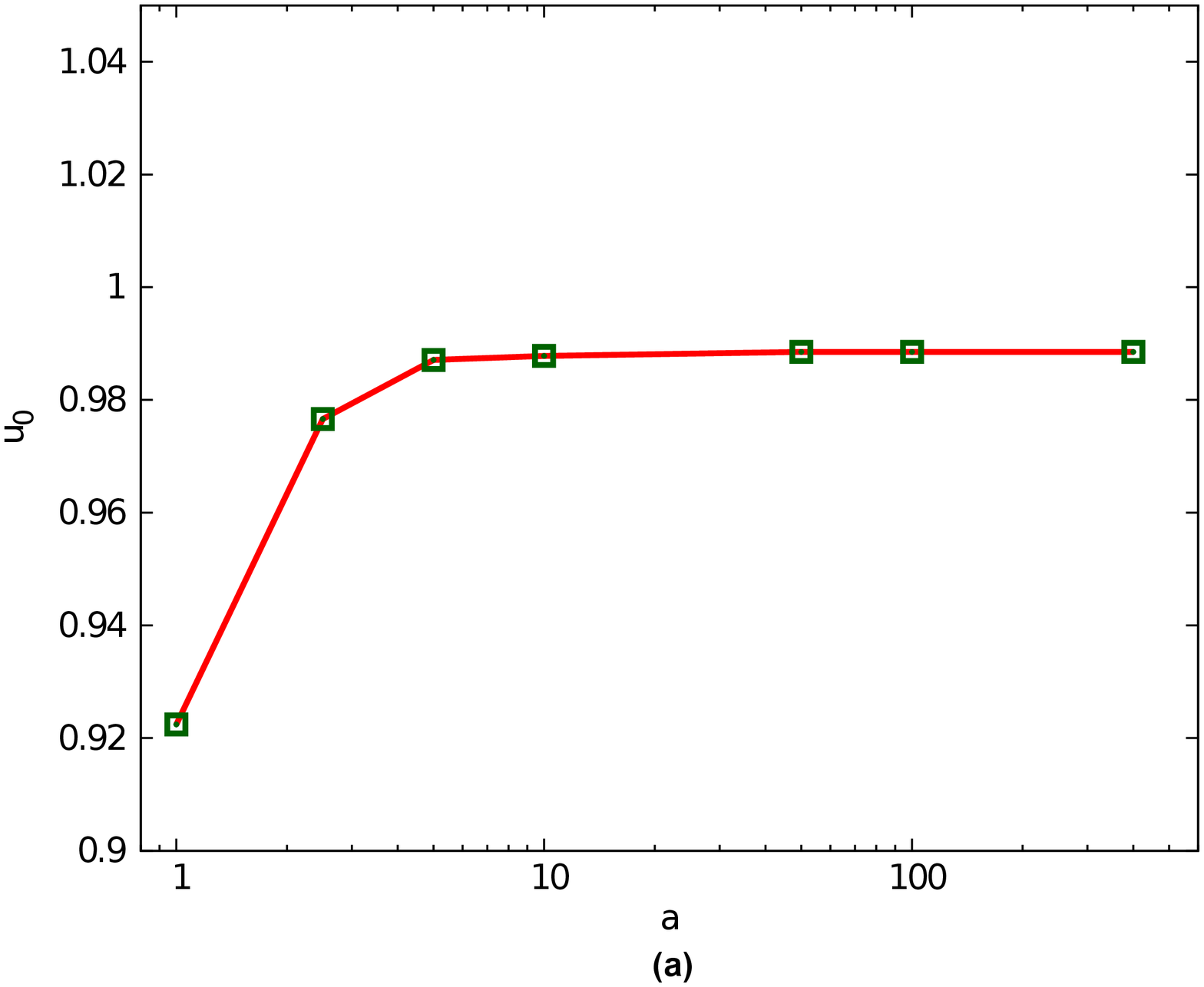}
\includegraphics[width=8cm]{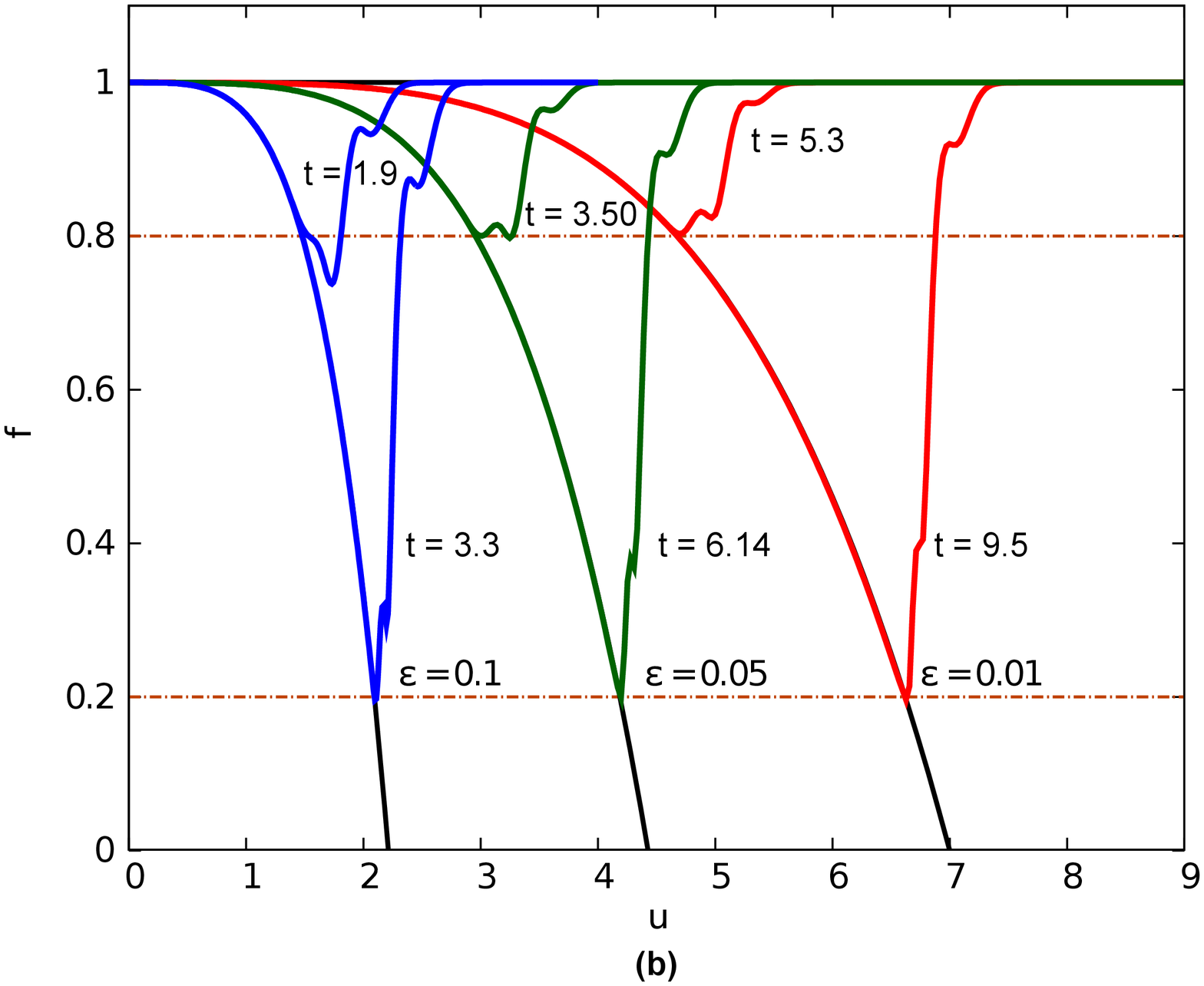}
\end{center}
\caption{The dependence of $u_0$ on $a$ for $\epsilon = 0.5$(Fig. (a)) and $f$ for $a = 10$ but different $\epsilon$(Fig. (b)). At $a\gtrsim5.0$, $u_0$ is almost independent of $a$. This fact and the scaling transformation in (\ref{equ:dilation}) allow us to conclude the relation in (\ref{equ:TPoincare}), which is verified in Fig. (b).
}\label{fig:aPoincare}
\end{figure} 

Let us first fix $\epsilon = 0.5$. As showed in Fig. \ref{fig:aPoincare}(a), $u_0$ only weakly depends on $a$ as $a\gtrsim5$. Using the scaling transformation in eq. (\ref{equ:dilation}) and neglecting such a weak dependence on $a$, one can conclude that as long as the condition \be\f{\epsilon}{a}\lesssim 0.1\label{equ:condPoincare}\ee holds, $u_0$ is (almost) independent of $a$ and  proportional to $\epsilon^{-\f{1}{2}}$, that is,
\be
u_0 = \f{u_0^{(0.5)}}{\sqrt{2\epsilon}} \simeq 0.699  \epsilon^{-\f{1}{2}},~~\mbox{or}, ~~T = \f{\sqrt{2\epsilon}}{\pi u_0^{(0.5)}}\simeq 0.455 \epsilon^{\f{1}{2}},\label{equ:TPoincare}
\ee
where $u_0^{(0.5)}$ is our numerical result with $(a,\epsilon)=(a^{(0.5)}, 0.5)$ and $a^{(0.5)}$ denotes the value of $a$ with fixe $\epsilon = 0.5$. The above relation $T\propto \epsilon^{\f{1}{2}}$ is also obtained by perturbative techniques in the limit $\f{\epsilon}{a}\ll1$ in Ref. \cite{Minwalla}.  Therefore, the perturbative techniques should be applicable when eq. (\ref{equ:condPoincare}) is satisfied. The condition (\ref{equ:condPoincare}) can be further rewritten in the following form
\be
\Delta t\equiv \f{1}{\sqrt{a}} \lesssim \f{0.14}{T}.\label{equ:narrowWave}
\ee
As a verification of (\ref{equ:TPoincare}),  Fig. \ref{fig:aPoincare}(b) shows  the metric function $f $ with $\epsilon = 0.1, 0.05$ and $0.01$ and fixed $a = 10$ at two different times. In region (I) they all agree with the AdS black hole metric with $u_0$ given by (\ref{equ:TPoincare}).

\begin{figure}
\begin{center}
\includegraphics[height=6.5cm]{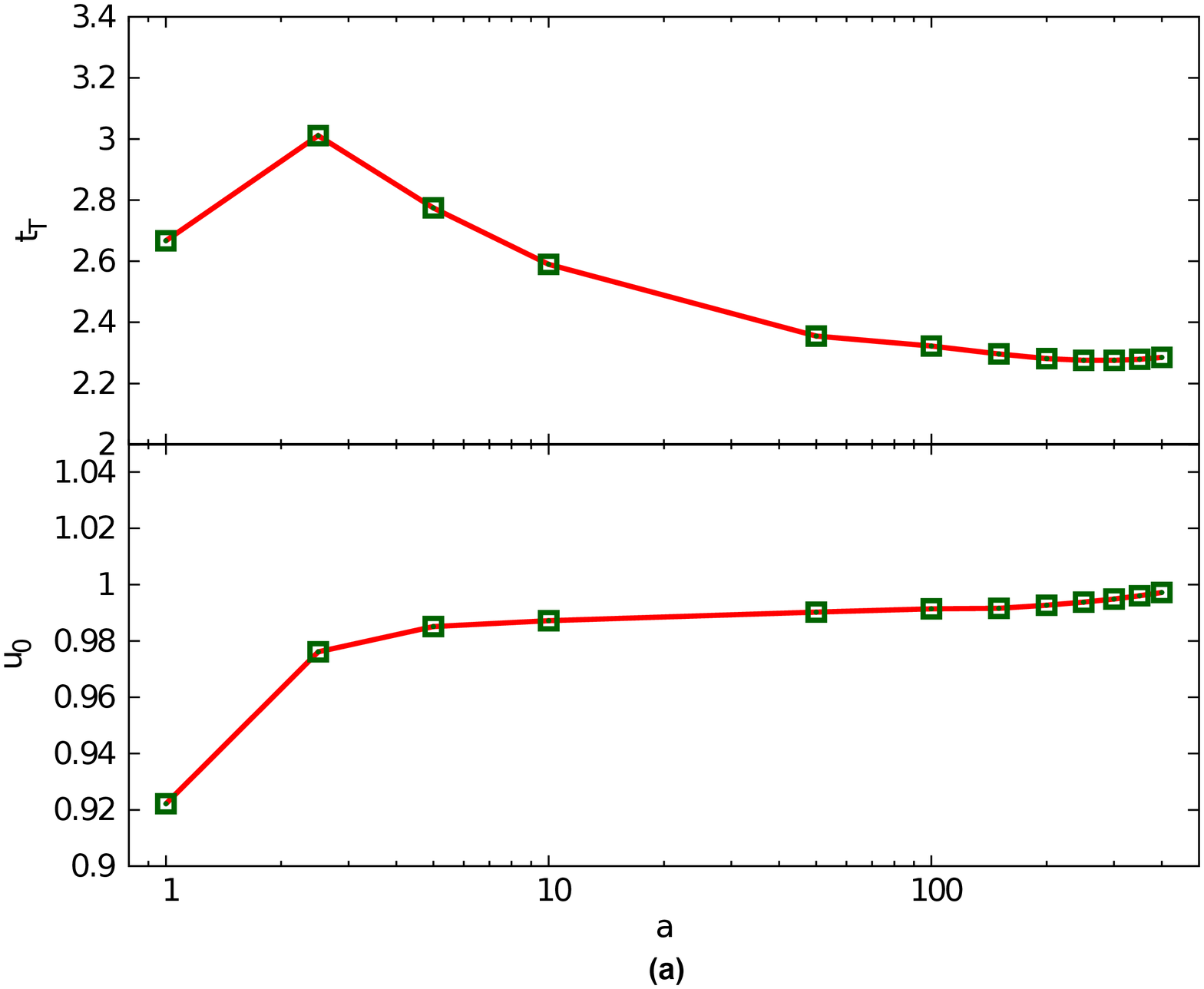}
\includegraphics[height=6.5cm]{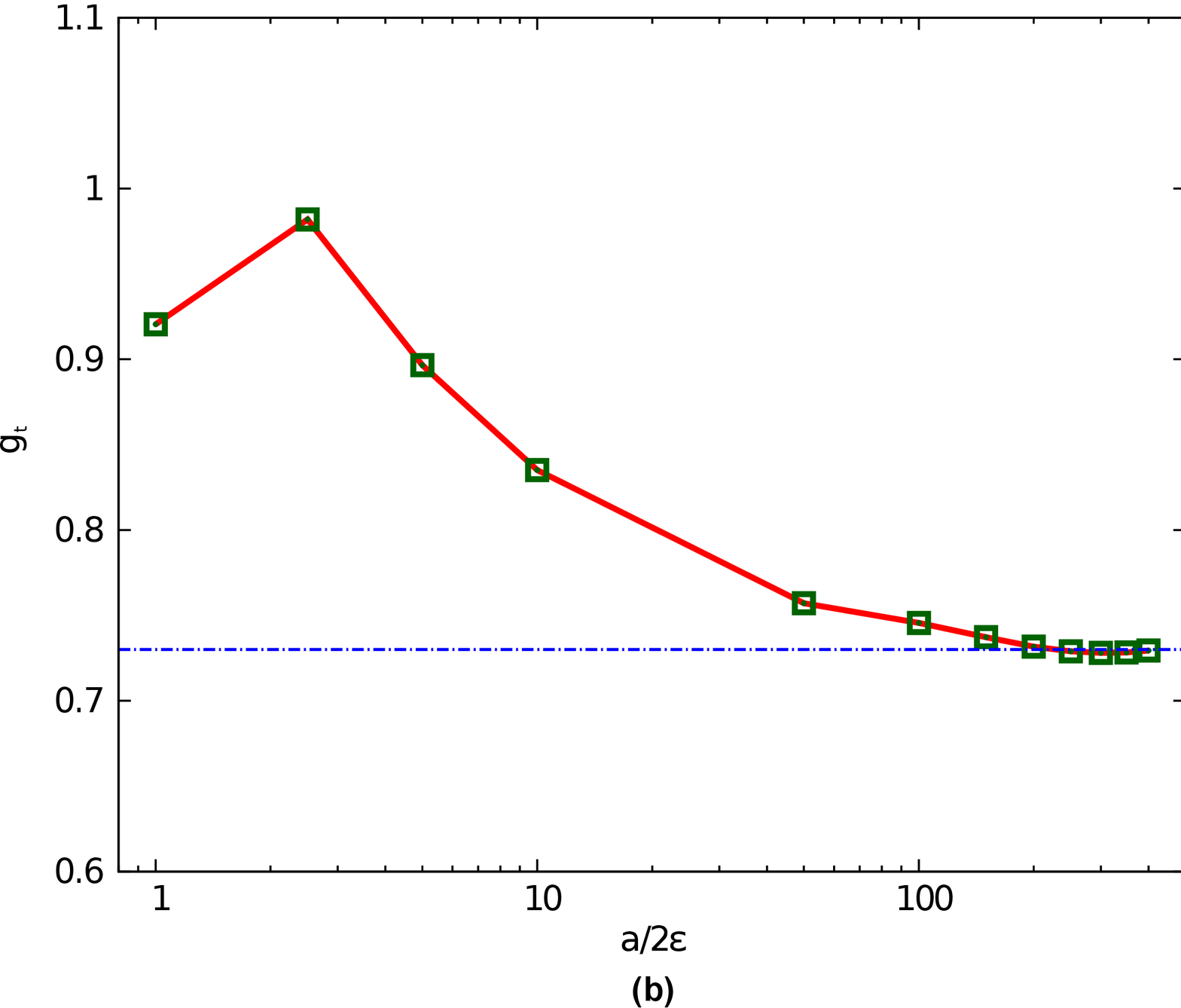}
\end{center}
\caption{$t_T$ and $u_0$ for $\epsilon = 0.5$(Fig. (a)) and $g_t$ as a function of $\f{a}{2\epsilon}$(Fig. (b)).
}\label{fig:g}
\end{figure} 

Fig.  \ref{fig:g}(a)  shows the thermalization time $t_{T}$ for $\epsilon = 0.5$ and different values of $a$. Again, the scaling transformation in (\ref{equ:dilation}) allows us to make conclusions for the cases with $(a, \epsilon)=(\lambda^2 a^{(0.5)}, 0.5 \lambda^2)\cong (a^{(0.5)}, 0.5)$, that is,
\be
t_T = \f{t_T^{(0.5)}}{\pi u_0^{(0.5)}}\f{1}{T}\equiv \f{g_t\left( \f{a}{2\epsilon} \right)}{T},\label{equ:narrowWavetT}
\ee
where $g_t\simeq 0.73$ for $\f{a}{2\epsilon}\gtrsim 200$ or, equivalently, $\Delta t \lesssim \f{0.02}{T}$ and $g_t\lesssim 1.0$ for $\f{\epsilon}{a}\lesssim 0.5$(see Fig. \ref{fig:g}(b)). Note that the thermalization time $t_T$ is physically different from the coordinate-dependent formation time of the black brane $t_{BH}\sim \Delta t$ in ingoing Eddington-Finkelstein coordinates reported in Ref. \cite{Minwalla}. We discuss this in detail in Appendix \ref{app:EddingtonFinkelstein}.

In the limit $a\to \infty$, one may make a simple estimate as follows: the source induces a three-peak wave in the bulk as showed in Fig. \ref{fig:e001a10}. The last(leftmost) peak does not contribute so much to the bending of the spacetime but does determine the location of the turning point between (I) and (II). Since the leftmost piece of this peak propagates in an approximate AdS black hole background, the thermalization time $t_T$ may be estimated by
\be
\Delta t_T \simeq \int_{0}^{u_{min}}\f{du}{f_{bh}} = \f{1}{4} u_{0} \left( 2 \arctan\f{u_{min}}{u_{0}} + \log \f{u_0+u_{min}}{u_0-u_{min}} \right)\simeq  2.02 u_0\simeq\f{0.64}{T}.\label{equ:narrowWavetTest}
\ee
At worst, this is the lower bound for $t_T$, which guarantees that the system thermalizes without violating causality. We take this as another justification for defining $t_T$ by eq. (\ref{equ:tT}). Taking $a = 400$, our numerical calculation gives $t_T\simeq 2.29$ and the estimated thermalization time is given by  $(\Delta t_T + 2 \Delta t )\simeq$ 2.12.

%Therefore, $t_T\sim\f{1}{T}$ indicates that the system achieves thermalization in a time scaling that saturates the causality bound.

%In Fig. \ref{fig:u0Poincare}, we show the results with $a=10$ and different $\epsilon$ and our numerical results fit the asymptotic $AdS$ black hole metric with $u_0$ given by eq. (\ref{equ:TPoincare}) very well.

%
%\begin{figure}
%\begin{center}
%\includegraphics[height=6cm]{tPoincare}
%\includegraphics[height=6cm]{tEPoincare}
%\end{center}
%\caption{The late time location of the scalar shell. Left panel: $f$ with $a = 100, 400$ and $500$ at different times. Right panel: energy density($f^2(V^2+P^2)$) for $a = 100$ and $500$ at t = 0.5 and 2.0. Here $\epsilon=0.5$.}\label{fig:tPoincare}
%\end{figure} 
%
\subsection{Broad waves($\Delta t \gtrsim \f{1}{T}$):  $\f{\epsilon}{a}\gtrsim1$}

In this subsection we discuss solutions with $\f{\epsilon}{a}\gtrsim 1$, or, $\Delta t\gtrsim \f{1}{T}$ by eq. (\ref{equ:narrowWave}). As illustrated in Fig. \ref{fig:2stagesThermalization}(a), the energy is injected into the CFT vacuum by the two pulses of $\dot{\phi}$. Based on the discussion in the previous subsection, the system is expected to achieve thermalization in a time scale  $t_T \sim \f{1}{T}$. Therefore, if $\Delta t \gtrsim \f{1}{T}$, one can expect a two-stage thermalization. Similarly, if $\dot{\phi}_0$ is periodic with period $\Delta t \gtrsim \f{1}{T}$, a multi-stage thermalization, as illustrated in Fig. \ref{fig:2stagesThermalization}(b), may also be expected.  We shall show how the results on the gravity side live up to such an expectation. 

\begin{figure}
\begin{center}
\includegraphics[width=8cm]{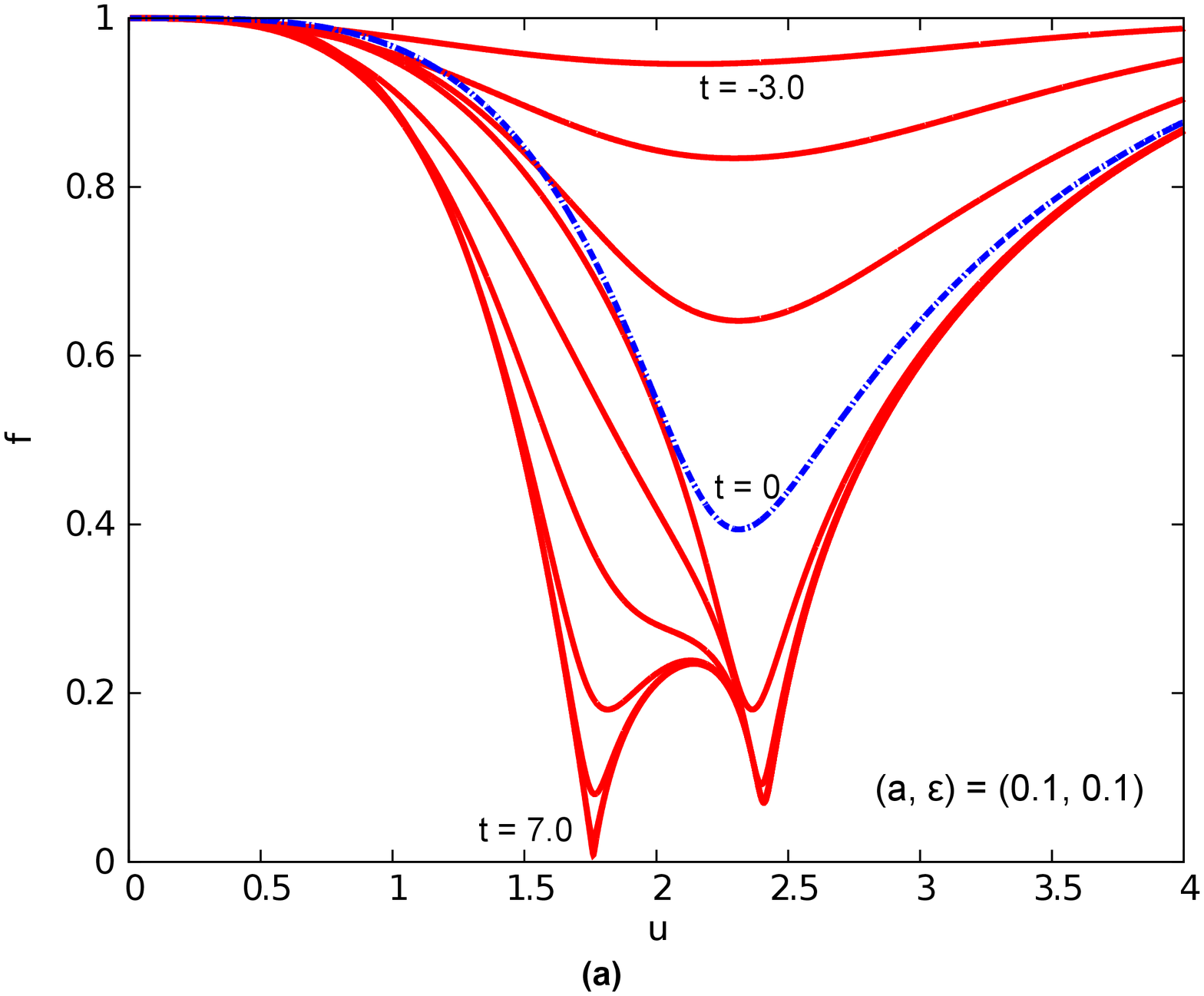}
\includegraphics[width=8cm]{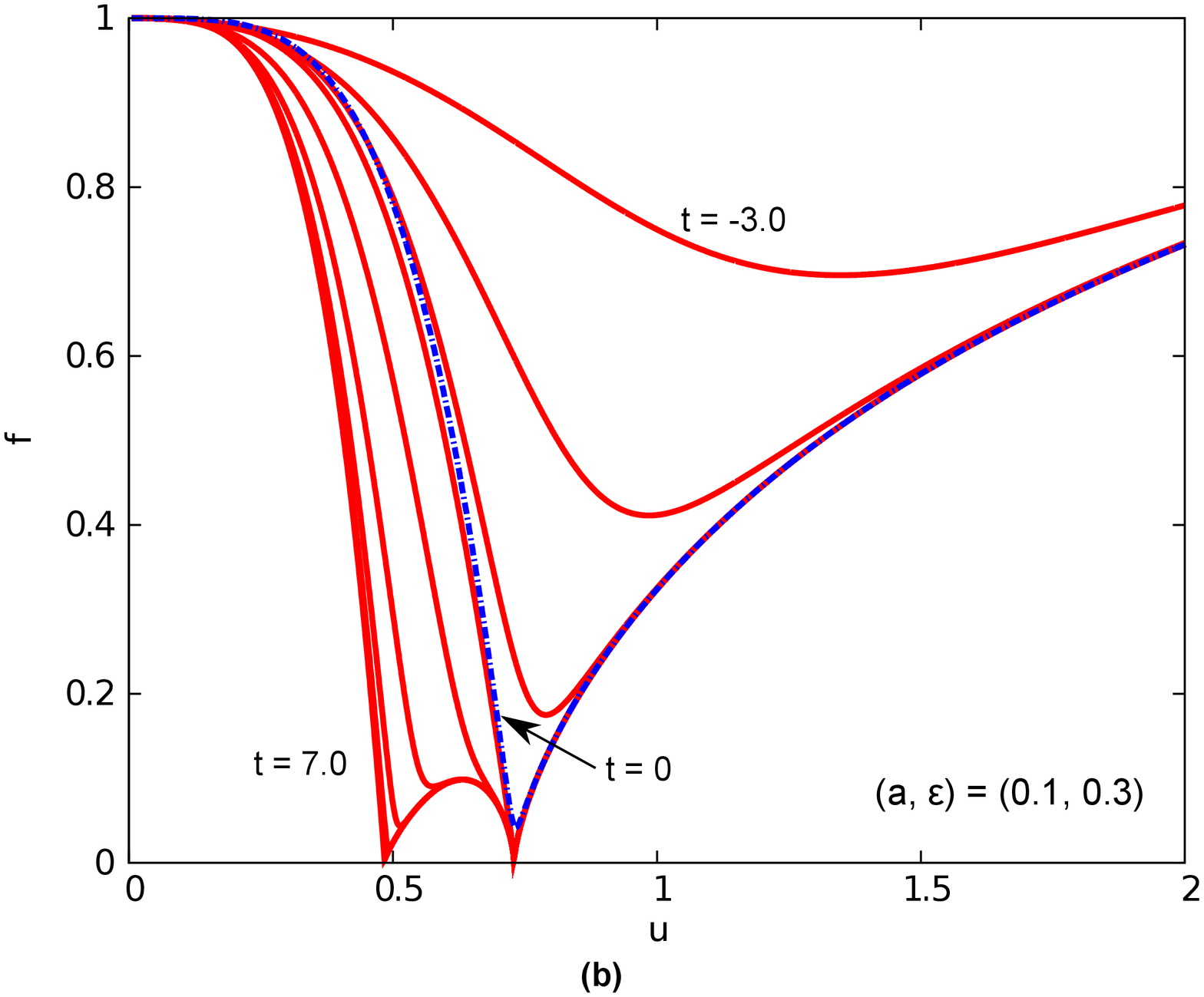}
\end{center}
\caption{Single-collapse(Fig. (a)) and double-collapse(Fig. (b)) solutions. In both figures we show $f$ as a function of $u$ from $t = -3.0$(top right) to 7.0(bottom left) with time steps equal to 1. In Fig. (a), $f$ at $t=0$ is far from that of the AdS black hole metric. In Fig. (b), $f$ at $t = 0$ is that of the AdS black hole metric at $u\leq u_{min}= 0.98 u_0$.
}\label{fig:twostageCollapse}
\end{figure}

In the cases with $\f{\epsilon}{a}\gtrsim 1$, it is more convenient to study solutions with $a$ fixed at a smaller value. We choose $a = 0.1$($\Delta t = 3.16$). Solutions with $1.0\geq \epsilon\geq 0.05$ are obtained by numerically solving (\ref{equ:eom}). We observe a transition from a single collapse to a double collapse in those solutions at $\epsilon \simeq (0.2-0.3)
$\footnote{Here, the first collapse at $t=0$ is determined by the following requirements: (a) $f(0,u)\geq f_{min}$ for all $u$, (b) $f(0,u_{min}) = f_{min}$ and (c) $f$ is that of the AdS black hole metric at $u\leq u_{min}$. The exact transition value of $\epsilon$ depends on the choice of $f_{min}$ or, equivalently, the scale $l$ for $\left< W(l) \right>$. For the solution with $\f{\epsilon}{a}=3.0$ we have $l\simeq\f{0.7}{T}$ or, equivalently, $u_{min} \simeq 0.98 u_0$ for the first collapse. At $t = 0.61$, min$\{f\}\simeq 0.01$. However, $f$ is not that of the AdS black hole metric near the boundary. Therefore, we do not interpret it as being in thermal equilibrium.}. 
Fig. \ref{fig:twostageCollapse} shows the metric function $f$ of a single-collapse solution($\epsilon = 0.1$) and a double-collapse solution($\epsilon=0.3$). In the single-collapse solution the scalar field collapses once at $u_0 = 1.76$ and $t = 6.61\sim 2\Delta t$. In contrast, in the double-collapse solution the scalar field collapses twice respectively at $u\simeq u_L = 0.73$ and $t\simeq 0$
\footnote{Here, $u_L$ is obtained by the least-squared fit of $f_{bh}$ with $u_0 = u_L$ to our numerical results for $u<u_{min}$.} 
and at $u \simeq u_H=0.48$ and $t=5.47\sim 2\Delta t$. For the first collapse, $\f{1}{T} = \pi u_L = 2.29 < \Delta t$ while for the single-collapse solution $2 \Delta t>\f{1}{T} = 5.53 > \Delta t$. The scalar fields induced by the source with $\epsilon \gtrsim 0.3$ all undergo such a double collapse. 
\begin{figure}
\begin{center}
\includegraphics[width=5cm]{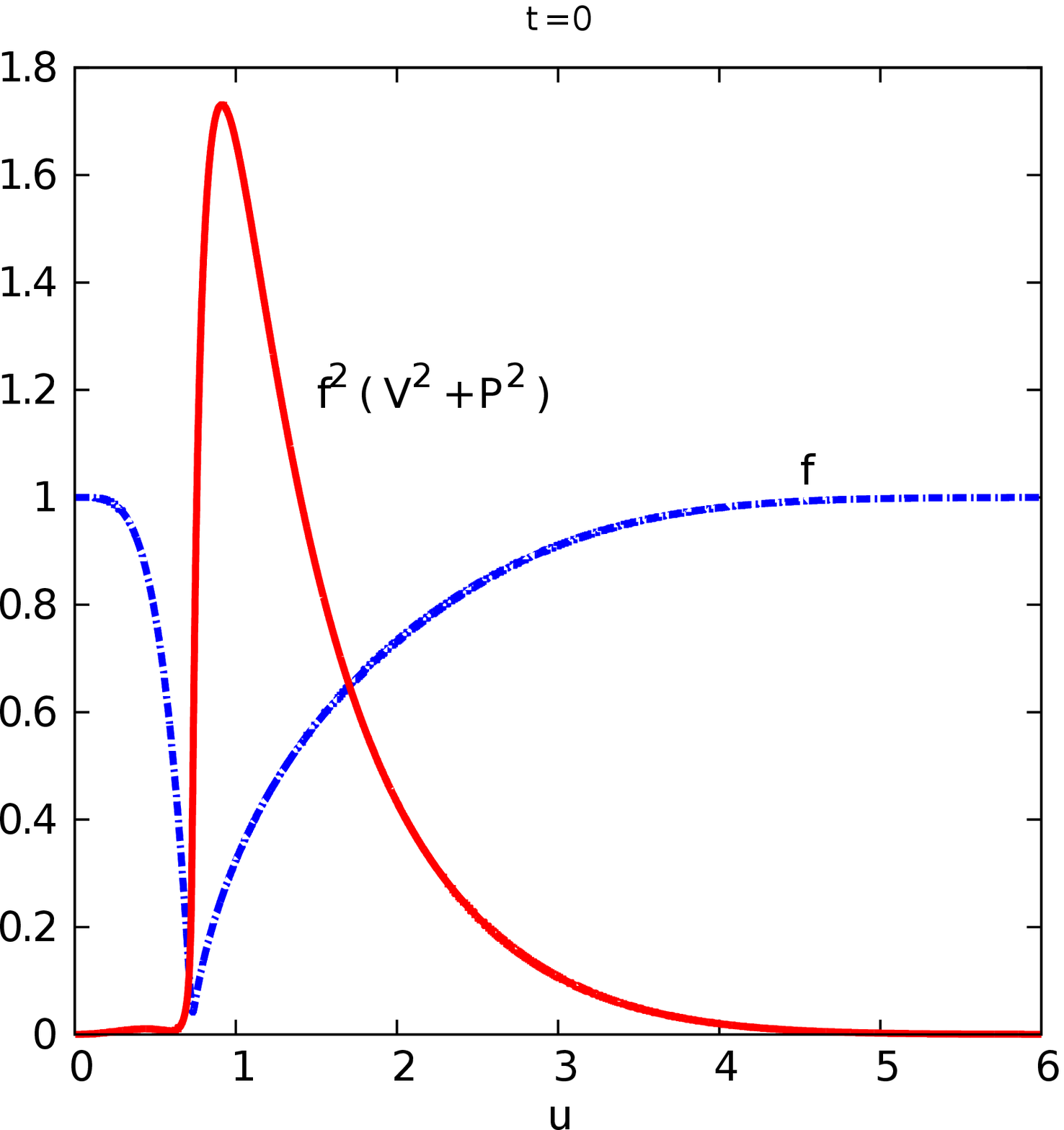}
\includegraphics[width=5cm]{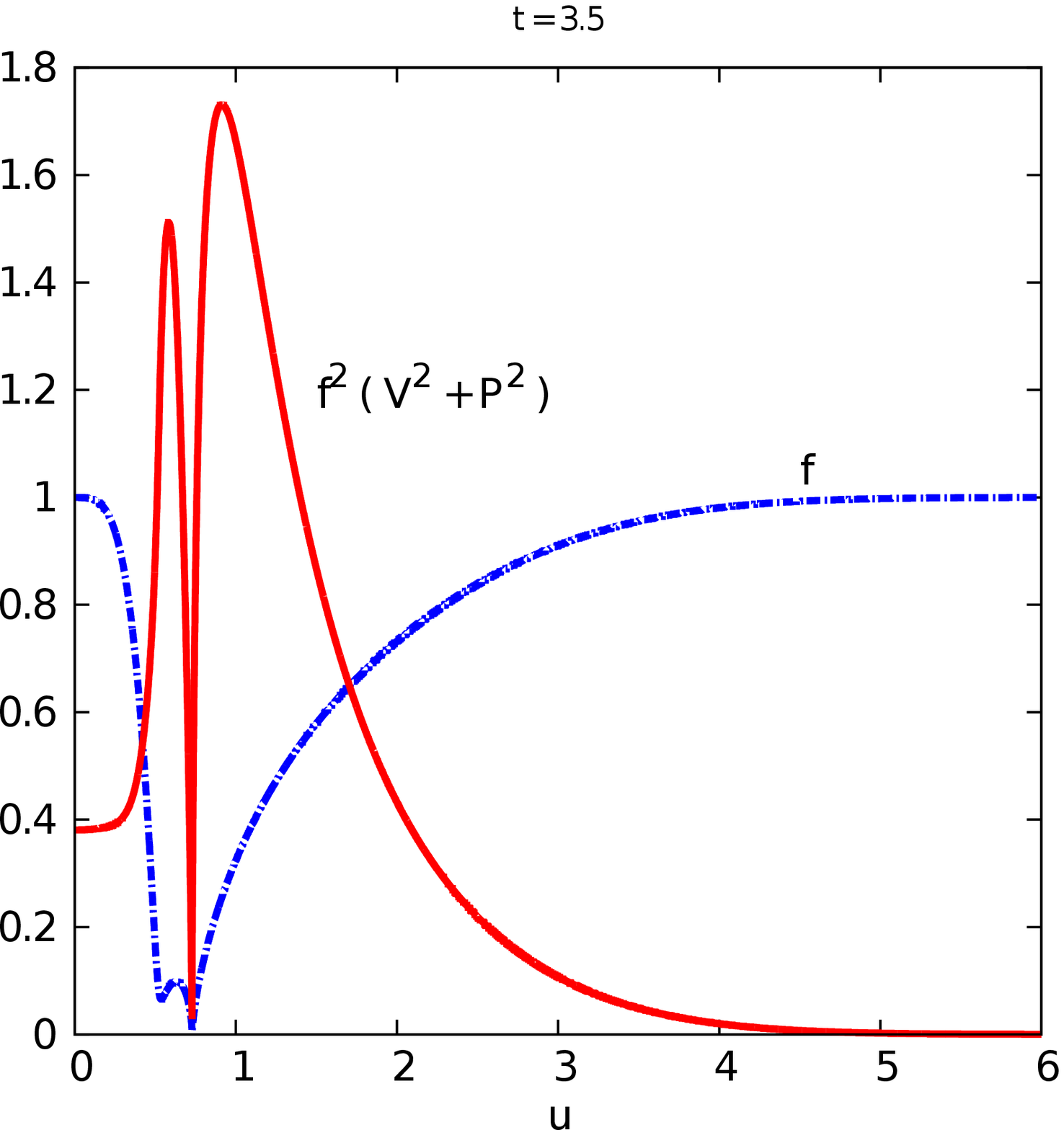}
\includegraphics[width=5cm]{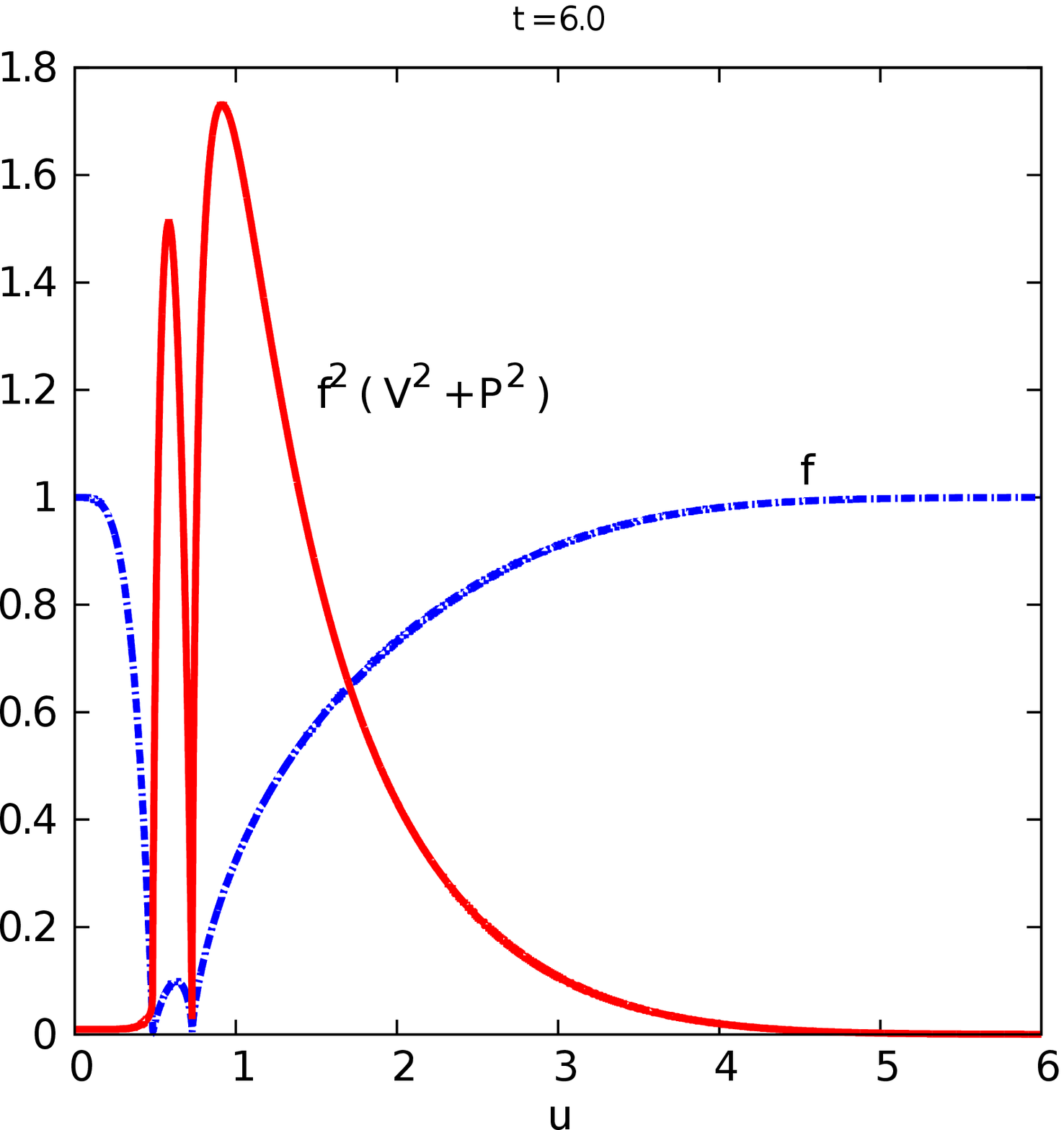}
\includegraphics[width=5cm]{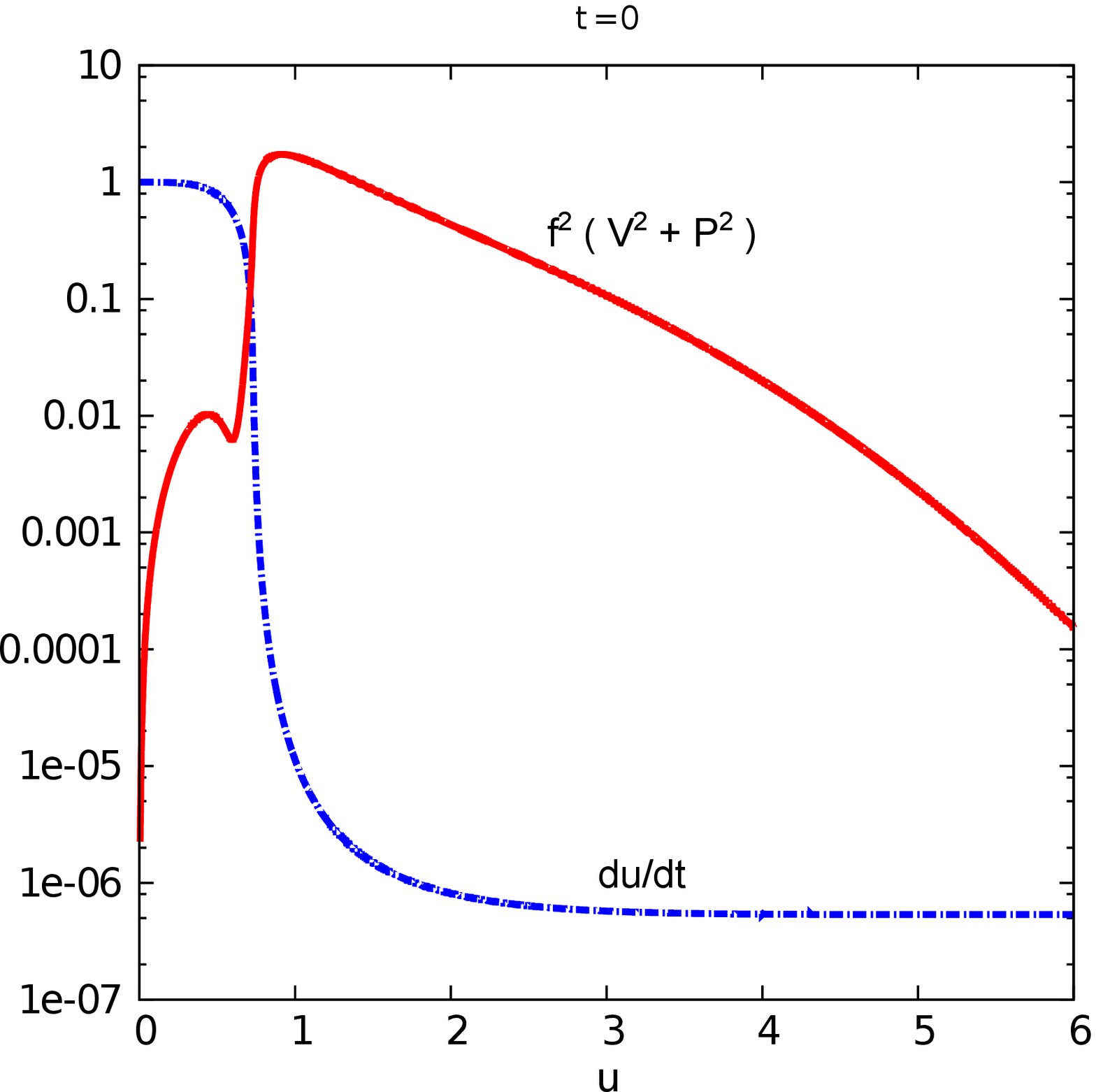}
\includegraphics[width=5cm]{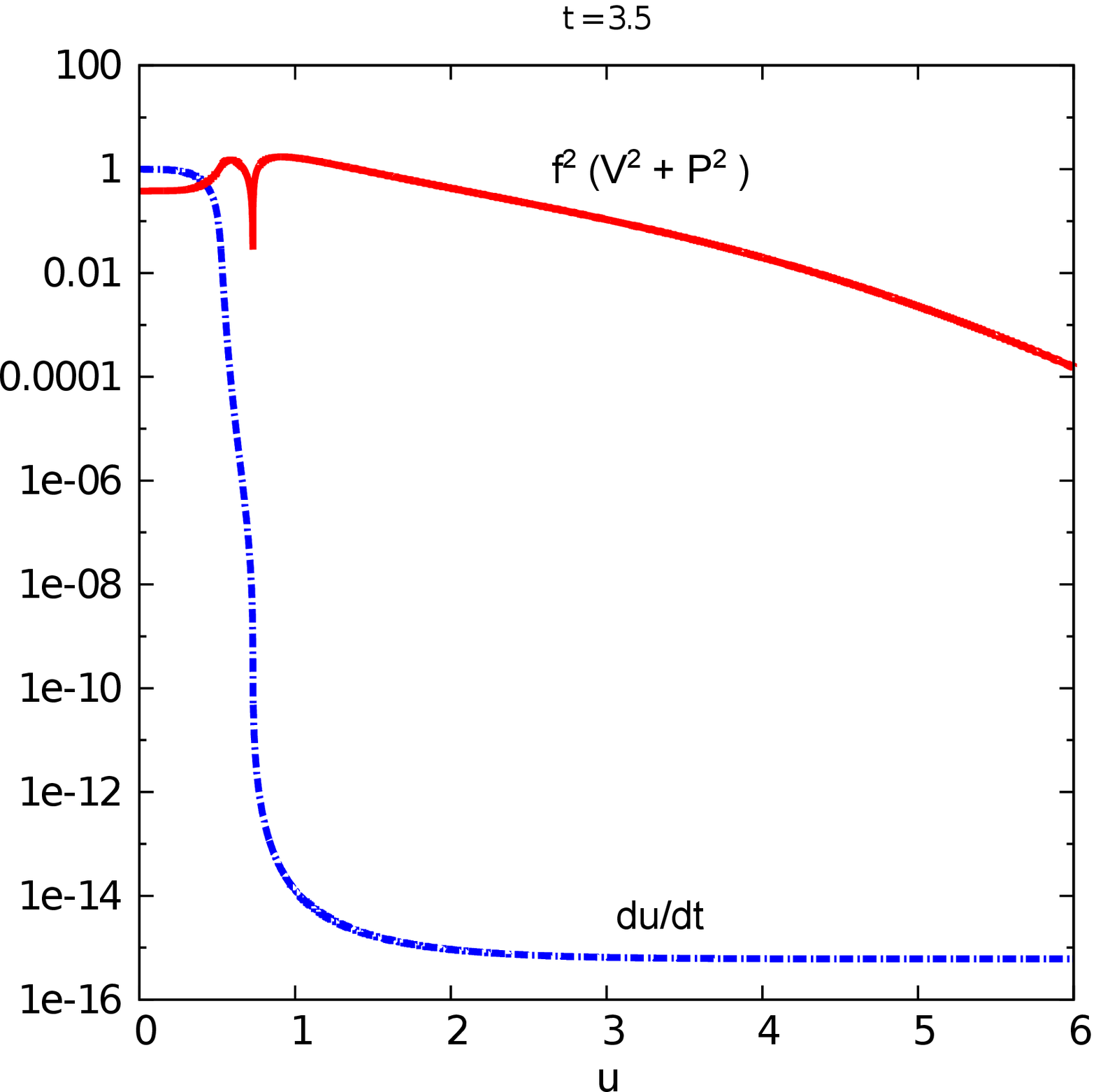}
\includegraphics[width=5cm]{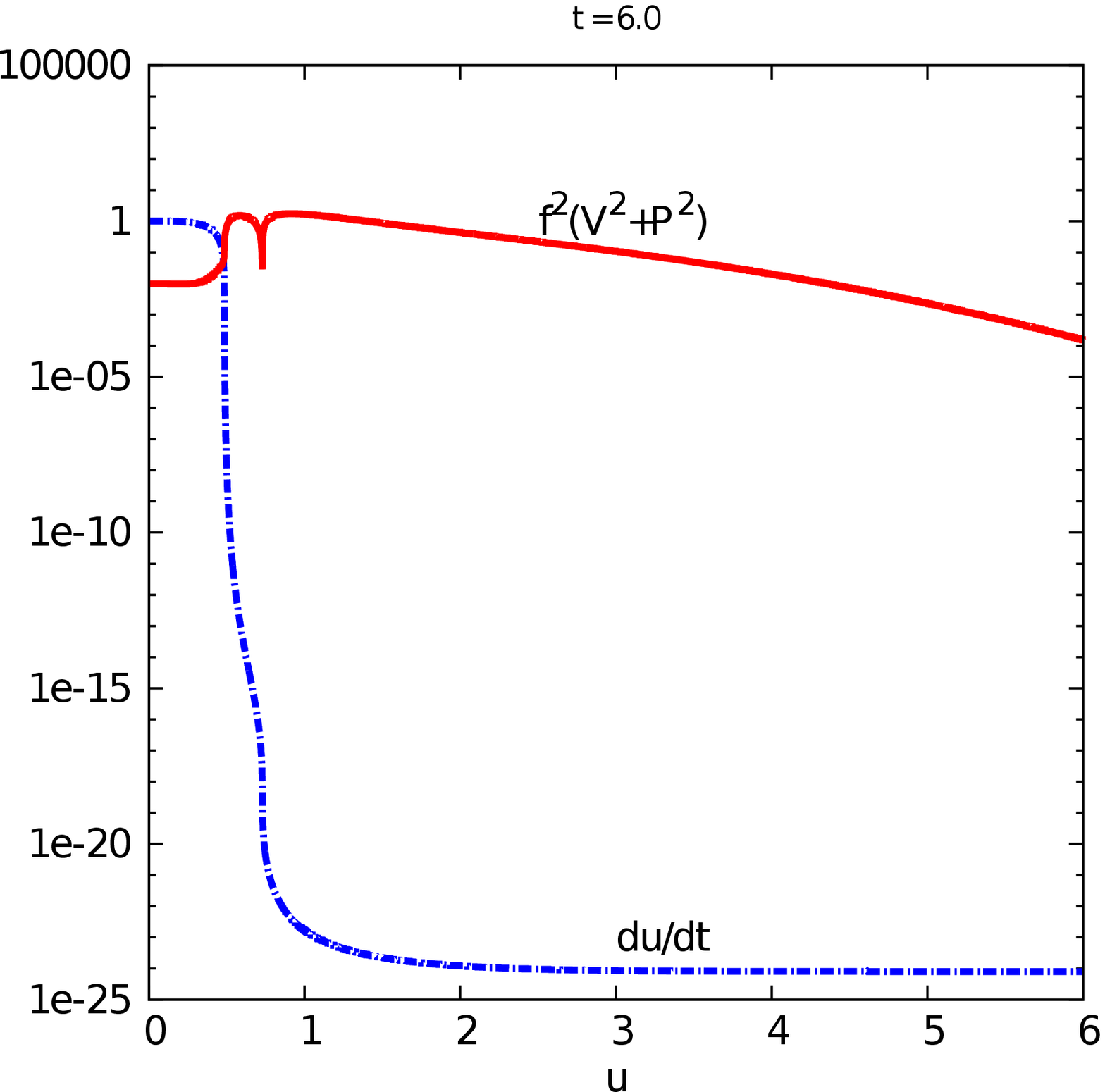}
\end{center}
\caption{The energy density of the scalar field($f^2(V^2+P^2)$), $f$ and the speed of light($du/dt=f e^{-\delta}$) for the double-collapse solution with $(a, \epsilon) = (0.1, 0.3)$.
}\label{fig:Ee03a01}
\end{figure}

Let us understand better the double-collapse solution. In Fig. \ref{fig:Ee03a01} we plot the energy density of the scalar field, $f$ and the speed of lightlike geodesics($du/dt = f e^{-\delta}$) of the solution with $\epsilon = 0.3$. At $t\geq 5.47$, two peaks are observed in the energy density. They correspond to the two collapses at $u\sim u_L$ and $u\sim u_H$. Also the speed of light drops dramatically in the bulk region $u\gtrsim u_H$. The peak at $u\sim u_L$ is resulted from the first collapse at $t\simeq0$ when the speed of light starts to drop below $10^{-6}$ at $u\gtrsim u_L$. From $t = 0$ to $5.47$, the scalar field mainly accretes at $u\sim u_H$, which only significantly modifies $f$ at $u\lesssim u_L$ and results in the second peak of the energy density.
\begin{figure}
\begin{center}
\includegraphics[width=8cm]{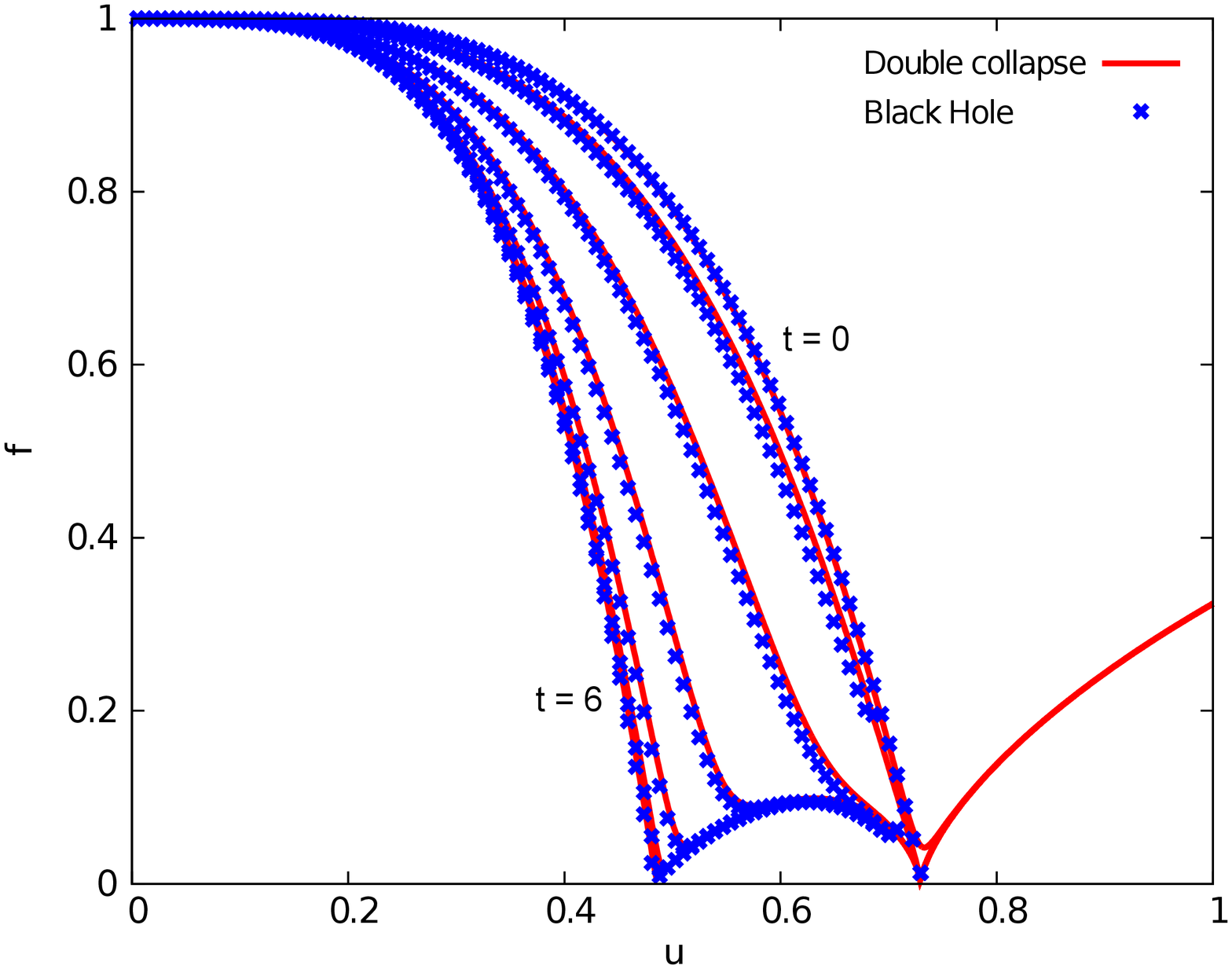}
\includegraphics[width=8cm]{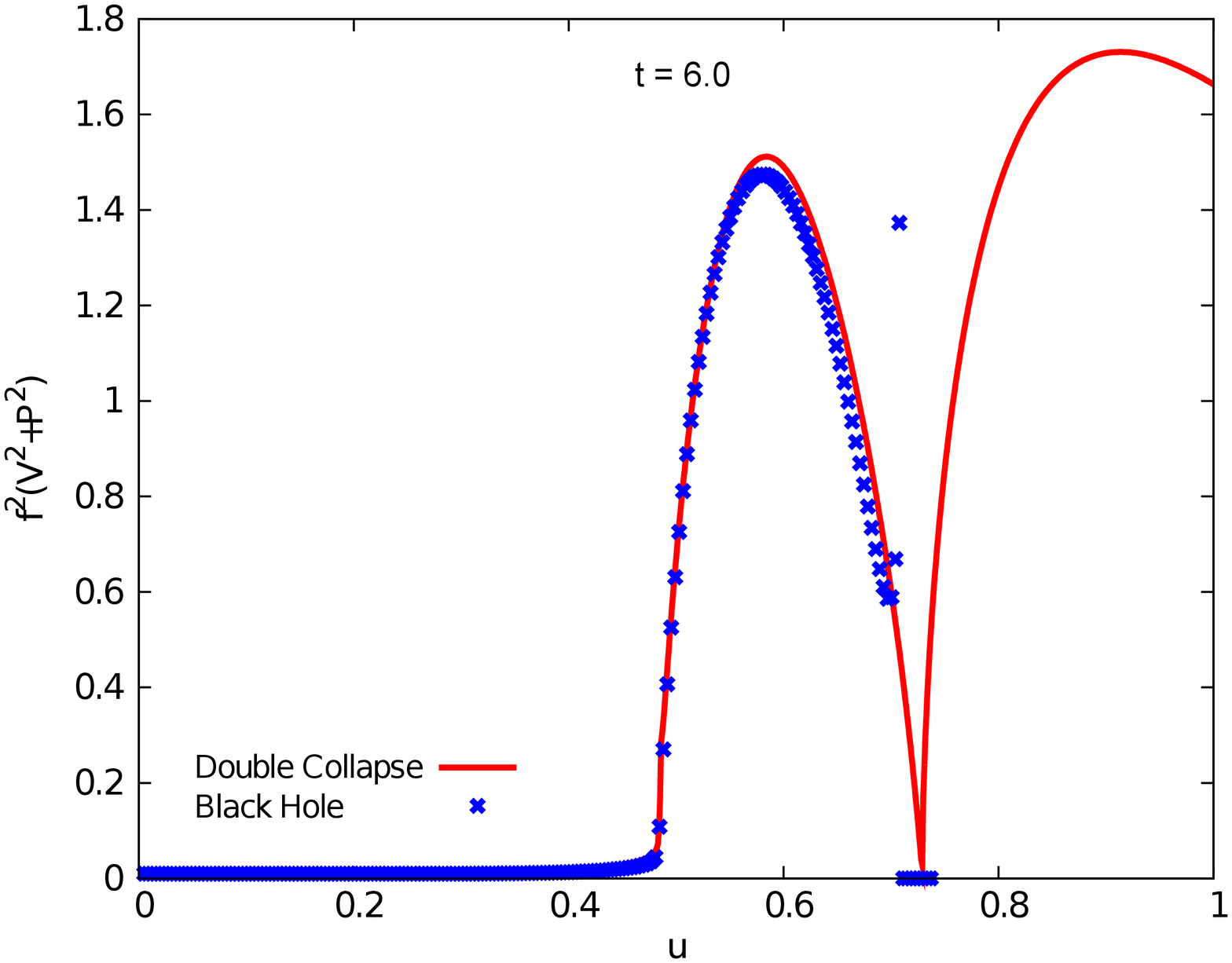}
\end{center}
\caption{Collapse of a massless scalar field on the AdS black hole background. In both figures, the dotted curves(Black Hole) show the results of turning on the scalar source at $t\geq0$ on the AdS black hole background. The solid curves(Double Collapse) show the results of the double-collapse solution with $(a, \epsilon) = (0.1, 0.3)$. In the left figure,  $f$ is showed as a function of $u$ from $t = 0$(top right) to 6.0(bottom left) with time steps equal to 1.
}\label{fig:responsebh}
\end{figure}

As another support for regarding the state at $t=0$ as being in (approximate) thermal equilibrium, we study the response of static plasma with $T = T_L$ to the scalar source which is turned on only after $t=0$. On the AdS side, we solve eq. (\ref{equ:eom}) using the AdS black hole metric in (\ref{equ:adsbh}) as the initial conditions. Here, we only discuss the solution with $T = T_L = 0.73$ and $(a,\epsilon)=(0.1,0.3)$.  Fig. \ref{fig:responsebh} shows the metric function $f$ at different times and the energy density at $t = 6.0$ of this solution. Their small difference from those of the double-collapse solution with $(a,\epsilon)=(0.1,0.3)$ justifies that an intermediate thermal equilibration is established at $t\simeq0$. 

\begin{figure}
\begin{center}
\includegraphics[width=8cm]{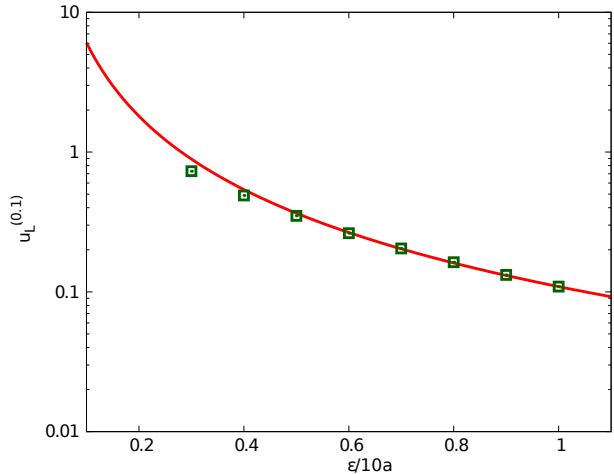}
\end{center}
\caption{$u^{(0.1)}_L$ as a function of $\epsilon/10a$.
}\label{fig:uL}
\end{figure}

Boundary conditions with $\f{\epsilon}{a}\gtrsim 3$ all give such double-collapse solutions\footnote{We have not calculated solutions with even larger $\f{\epsilon}{a}>10$ because it is numerically too time-consuming. However, based on the physical argument given in the paper, it is reasonable to expect that the conclusion should also apply to the cases with $\f{\epsilon}{a} > 10$.} by the scaling transformation in (\ref{equ:dilation}).  In general we conclude that the criteria for the system to achieve thermalization induced by the first pulse of $\dot{\phi}$ at $t\simeq0$ is $\Delta t \gtrsim \f{1}{T_L}$. This is exactly what we expected on the CFT side. The intermediate thermalization temperature $T_L$ for arbitrary $(a, \epsilon)$ is given by
\be
T_L =\f{1}{\pi u_L}= \f{\sqrt{10 a}}{\pi u_L^{(0.1)}\left( \f{\epsilon}{10a} \right)},
\ee
where $u_L^{(0.1)}\left( \f{\epsilon}{10a} \right)$, showed in Fig. \ref{fig:uL}, is our numerical results with $a=0.1$. At larger $\epsilon$, $u_L^{(0.1)}$ scales according to the following power law
\be
u_L^{(0.1)}(\epsilon) \simeq 0.109 \epsilon^{-\beta}\label{equ:uL}
\ee
with $\beta \simeq 1.74$. The second collapse(the heating-up process) can also be studied by turning on the source at $t\geq0$ and starting from the initial conditions in (\ref{equ:adsbh}) with $u_0 =u_L = \f{1}{\pi T_L}$, as we did for the case  with $(a, \epsilon) = (0.1, 0.3)$.
\begin{figure}
\begin{center}
\includegraphics[height=6cm]{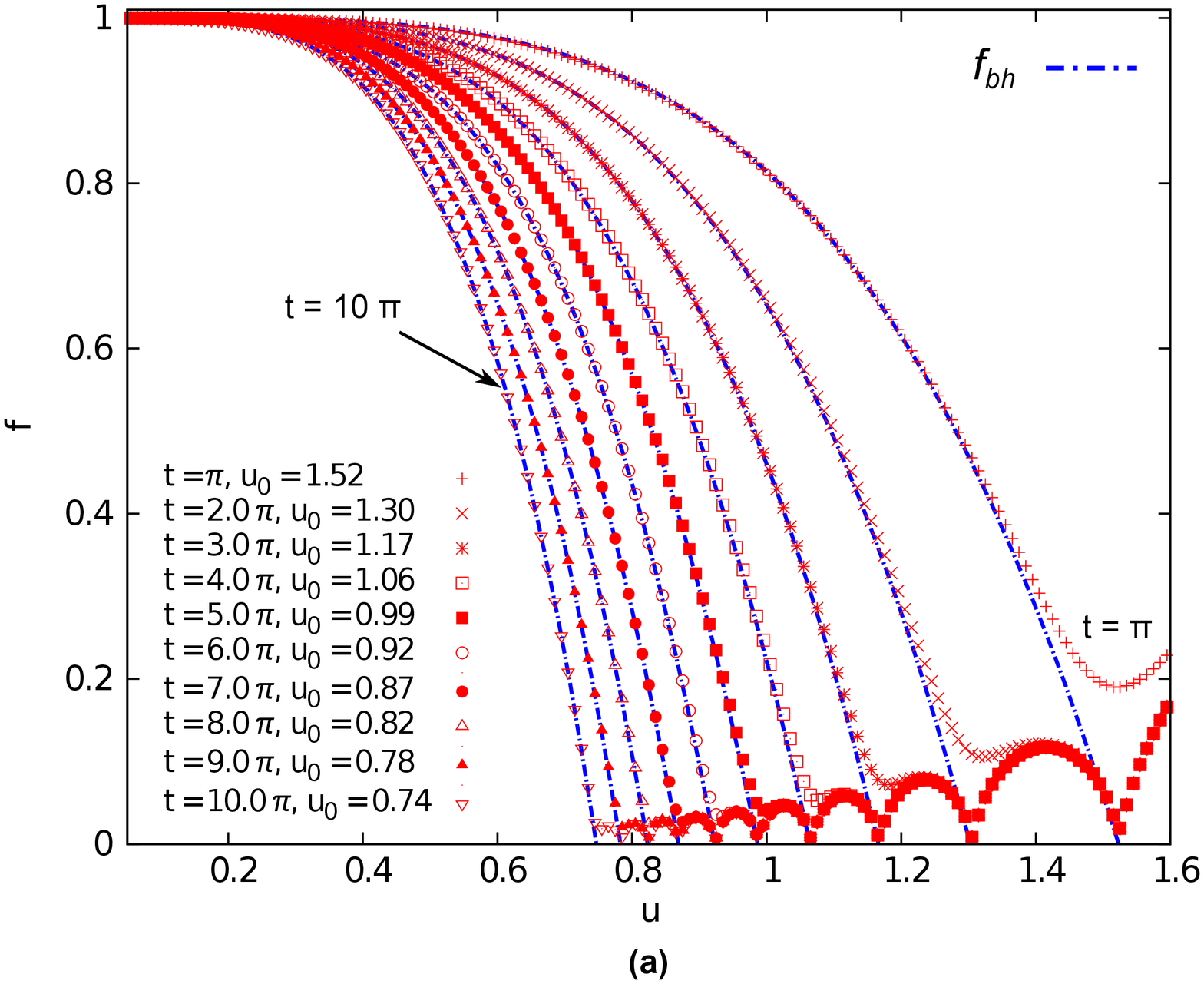}
\includegraphics[height=6cm]{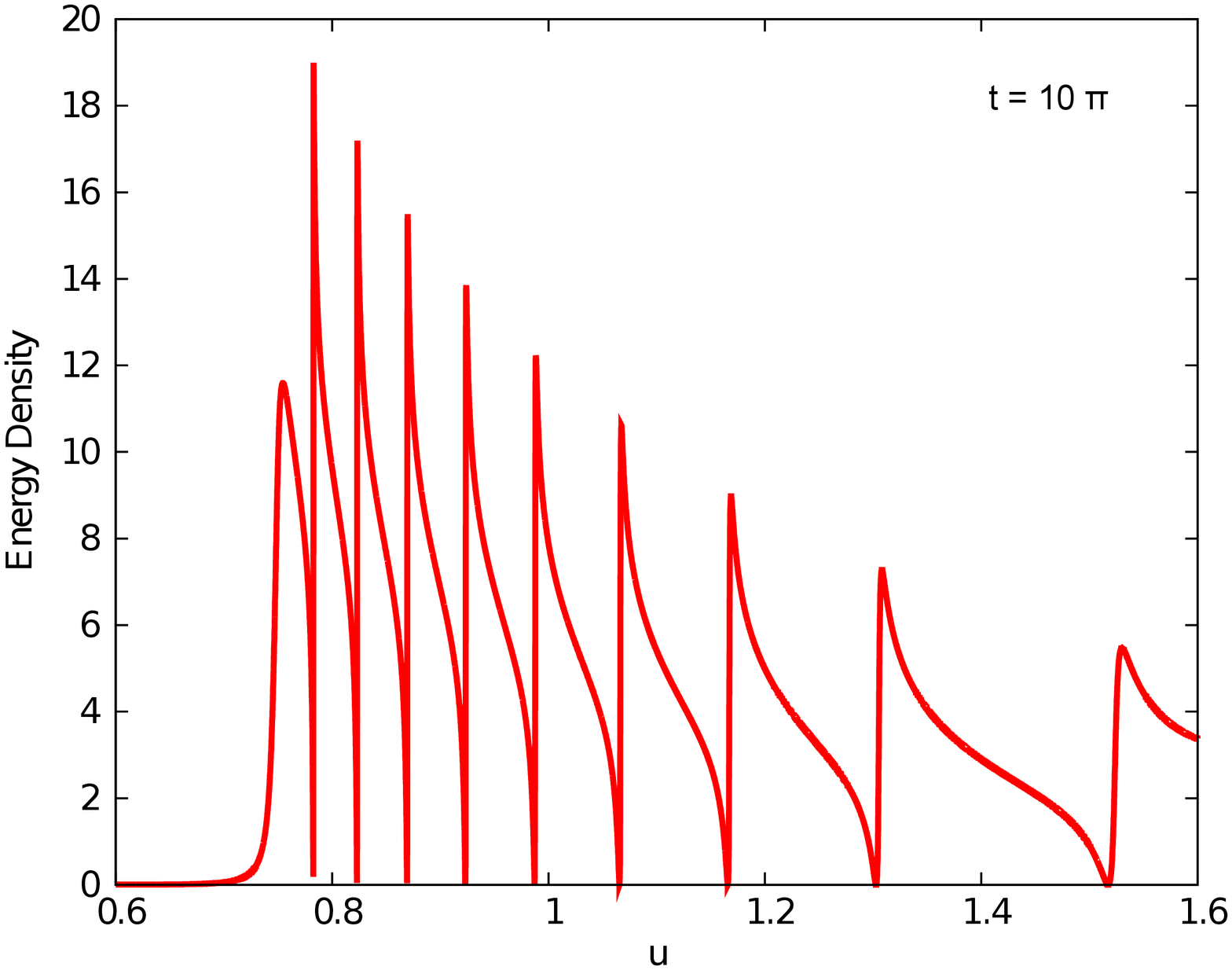}
\end{center}
\caption{A multiple-collapse solution. Fig. (a) shows $f$ and $f_{bh}$ at different times from $t=\pi$(right) to $t=10\pi$(left) with time intervals equal to $\pi$. Here $u_0$ in $f_{bh}$ is obtained by performing least-square fit of the AdS black hole metric to our results at $u\leq 0.505$ with $\sigma \sim 10^{-4}$. Fig. (b) shows the energy density of the scalar field ($f \left( V^2 + P^2 \right)$) at $t = 10 \pi$. Each spike here is the result of the collapse around $t = \pi, 2 \pi, \cdot\cdot\cdot, 10\pi$.
}\label{fig:sin}
\end{figure}

The qualitative conclusion above is independent of the source's shape. As a justification, Fig. \ref{fig:sin} shows a multiple-collapse solution. The periodic source is given by
\be
\dot{\phi}_0 = \epsilon \sin(t) \theta(t)\label{equ:sin}
\ee 
with $\epsilon=0.3$. Here, one naturally takes $\Delta t = \pi$. Fig. \ref{fig:sin}(a) shows $f$ at $t = \pi, 2\pi,\cdot\cdot\cdot, 10\pi$. At $t=\pi$, $f$ is different\footnote{Here, we define $u_{min}$ by $\left|\f{f_{min}}{f_{bh}(u_{min})}-1\right|=0.1$ with $f_{min}$ being our numerical result at $u=u_{min}$.} from that of the AdS black hole metric at $u\gtrsim u_{min}= 0.93 u_0$ and $\f{1}{T} = \pi u_0 = 1.52 \pi > \Delta t$. In contrast, at $t = 10\pi$, $f$ is that of the AdS black hole metric at $u\lesssim 0.99 u_0$ and $\f{1}{T} = 0.74 \pi$.  The energy density of the scalar field at $t=10 \pi$ is showed in Fig. \ref{fig:sin}(b). Each spike in the figure is the result of the collapse around $t = \pi, 2 \pi, \cdot\cdot\cdot, 10\pi$. Therefore, the parametric criteria for a system driven by a periodic source to achieve thermal equilibrium within each time interval $\Delta t$ is also $\Delta t \gtrsim \f{1}{T}$ with $T$ being the intermediate thermal equilibrium temperature.
\section{Discussions}\label{sec:discussion}
In this paper, we focus on the bulk metric resulted from gravitational collapse of a massless scalar field in the Poincare patch of $AdS_5$. Mostly, we aim to understand the typical thermalization time scale $t_T$ of its CFT dual. We find that thermalization in such a strongly coupled system is rapid in the sense that \be t_T \simeq \f{O(1)}{T},\ee
where the coefficient of $O(1)$ depends on the scale $l\sim \f{1}{T}$ of nonlocal operators and on the boundary source's shape. Such a rapid thermalization time seems to be typical of the strongly coupled CFT\cite{CY01,Heller:2012km}.
%This parametric relation should be generally hold in such a system but the coefficient of $O(1)$ depends the explicit shape of the source. 

We leave many unanswered but intriguing questions for future studies. There are still more details about such a thermalization process that can be understood only by evaluating non-local operators\cite{holographicThermal} or other relevant probes\cite{energylossandpT,Baier:2012a,Baier:2012b}. Moreover, in this paper we only study gravitational collapse of massless scalars. It is interesting to know what is the typical thermalization time $t_T$ in the cases of massive and tachyonic scalars, which are respectively dual to irrelevant and relevant operators in the boundary CFT\cite{Witten}.

%Besides,  CFT on $R\times S^3$ has different phases\cite{Witten}, which can be studied in the global patch of $AdS_5$\cite{Minwalla, Bizon01,Bizon02, Garfinkle01,Garfinkle02}. 
%
%%

\section*{Acknowledgements}
The author would like to thank R. Baier, Y. Kovchegov, P. Romatschke, S. Stricker and A. Vuorinen for reading this manuscript and providing illuminating comments. This work is supported by the Humboldt foundation through its Sofja Kovalevskaja program.

\appendix

\section{Equations of motion in Eddington-Finkelstein coordinates}\label{app:EddingtonFinkelstein}
In outgoing Eddington-Finkelstein coordinates the metric can be written in the following from
\be\label{equ:EddingtonFinkelstein}
ds^2=\f{1}{u^2}\left( -f e^{-2\delta} dv^2 + 2 e^{-\delta} dvdu + d\vec{x}^2 \right),
\ee
where $f$ and $\delta$ are functions of $v$ and $u$ only. We only need to solve the following equations of motion
\bea\label{equ:eomEdd}
&&\dot{f} = \f{4}{3} u V \left(f P + e^{\delta} V\right),~~\dot{P} = V',\nn\\
&&\delta' = \f{2}{3} u P^2,~~V'=\f{3}{2} \f{V}{u} - \f{1}{2} u^3 \left( \f{1}{u^3} f e^{-\delta} P\right)',
\eea
where  the derivatives with respect to $v$ and $u$ are denoted respectively by overdots and primes,
\be
P\equiv \phi'\mbox{~~and~~} V \equiv \dot{\phi}.
\ee
One constraint equation is given by
\be
f' = \f{2}{3} u P^2 f + \f{4}{u}\left(f-1\right).
\ee
%And the energy conservation equation is given by
%\be
%a_4' = \frac{1}{108} \left(576 p_4 \dot{p}_0 - 224 \dot{p}_0^3 \ddot{p}_0 - 90 \ddot{p}_0 p_0^{(3)}+3 \dot{p}_0 p_0^{(4)}\right)
%\ee
%In advanced Eddington-Finkelstein coordinates the metric can be written in the following from
%\be\label{equ:EddingtonFinkelsteinad}
%ds^2=\f{1}{u^2}\left( -f e^{-2\delta} dv^2 - 2 e^{-\delta} dvdu + d\vec{x}^2 \right),
%\ee
%equations of motion are given by replacing $e^{\pm\delta}$ with $-e^{\pm\delta}$ in (\ref{equ:eomEdd}).

\begin{figure}
\begin{center}
\includegraphics[width=7cm]{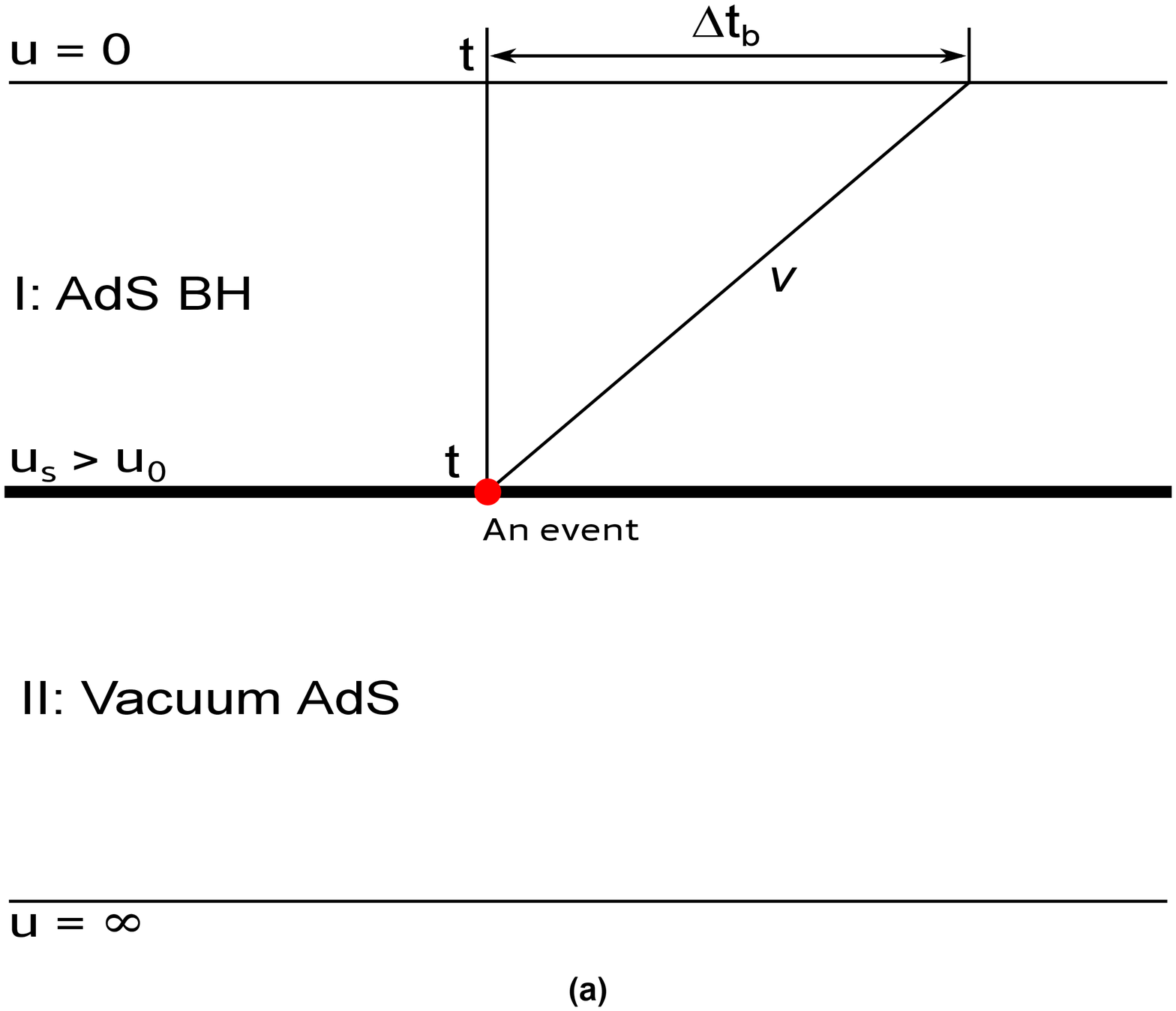}
\includegraphics[width=7cm]{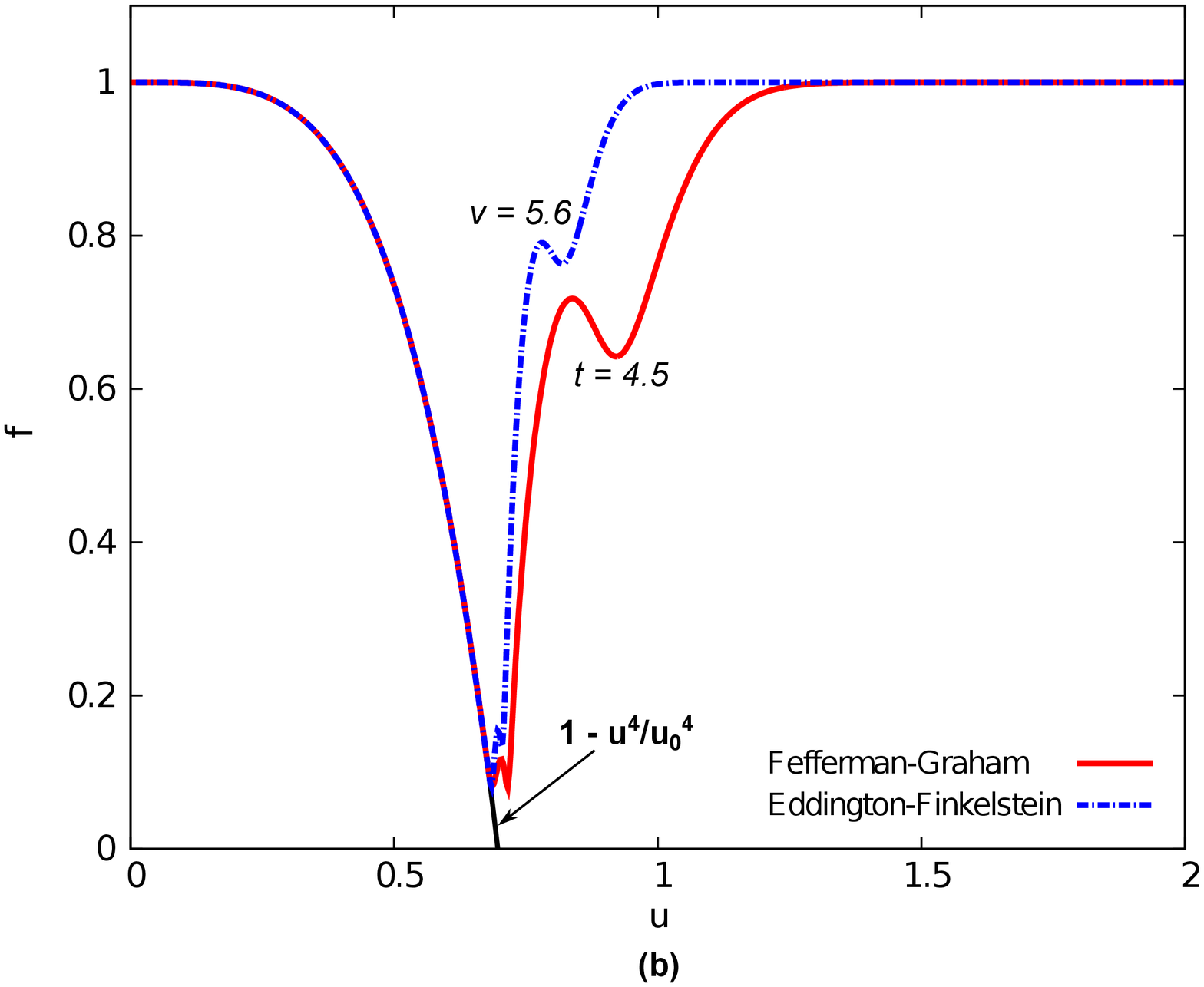}
\end{center}
\caption{Coordinate dependence of the boundary time. Fig. (a): An event occurs at $u=u_s$ and at $t$ in Schwarzschild coordinates. $\Delta t_b$ is the time difference from $t$ on the boundary in outgoing Eddington-Finkelstein coordinates. Fig. (b): Results of $f$ calculated in these two coordinate systems. Here $(a, \epsilon) = (10,1)$,  $u_s = u_{min}= 0.685$ and $u_0 = 0.697$. The event in this example is that $f$ drops to $f_{min}=0.07$ at $u_{min}$ for the first time.
}\label{fig:formationTime}
\end{figure} 

The simple form of the definition of $t_T$ in eq. (\ref{equ:tT}) is due to our choice of coordinates. If we naively define $t_T$ by the same definition in (\ref{equ:tT}) in both coordinate systems,  the thermalization time $t_T$ calculated in Eddington-Finkelstein coordinates in (\ref{equ:EddingtonFinkelstein}) is larger than that in Schwarzschild coordinates in (\ref{equ:FeffermanGraham}) with a time difference given by
\be
\Delta t_b = \int_0^{u_{min}} \f{du}{f} = \f{1}{4} u_{0} \left( 2 \arctan\f{u_{min}}{u_{0}} + \log \f{u_0+u_{min}}{u_0-u_{min}} \right),\label{equ:dtb}
\ee
where we have taken $f = 1 - u^4/u_0^4$. The reason for this is illustrated in Fig. \ref{fig:formationTime}(a).  Fig. \ref{fig:formationTime}(b) shows one example with $u_{min} = 0.685$ and $u_0 = 0.697$. In this case, eq. (\ref{equ:dtb}) gives\footnote{Here, we choose $f(t\text{ or }v,u_{min}) = f_{min}$ with $f_{min}=0.07$ for saving computation time.}
\be
\Delta t_b \simeq 1.10,
\ee
which is exactly what we find in our numerical results. Therefore, to get the same $t_T$ in Eddington-Finkelstein coordinates  one has to solve the spacelike geodesic equation (76) in Ref. \cite{holographicThermal}, which complicates our numerical calculations. Similarly, in ingoing Eddington-Finkelstein coordinates
\footnote{However, in ingoing Eddington-Finkelstein coordinates, say, $ds=\f{1}{u^2}\left( -f e^{-2\delta} dv^2 - 2 e^{-\delta} dvdu + d\vec{x}^2 \right)$ it is difficult to provide initial conditions for numerical simulations in our case.} 
such as that used in Ref. \cite{Minwalla} one can expect a shorter formation time of the black brane. 
%(v1)This makes it difficult to compare $t_T$ with the formation time of the black brane reported in another form of Eddington-Finkelstein coordinates in Ref. \cite{Minwalla}.

% but the knowledge about the bulk metric in the furthe $v$ is needed. For such a simplicity we will only show the results in Schwarzschild coordinates in this paper.
%
% to probe high momentum modes($k\sim \f{1}{l} \ll T$) in CFT by the Wilson loop  one only needs to know the metric of the submanifold at $u\lesssim l$ while for low momentum modes($k\sim \f{1}{l} \sim T$) all the information of the metric at $u\leq u_{min}\sim u_0$ is needed. 
% Since $u(0)$ is a monotonically increasing function of $l$(Fig. \ref{fig:wilsonloop}(a)), at $t_T$ the boundary system has achieved thermalization for th e wilson loop $\left<W(l)\right>$ at all scales $l\lesssim \f{1}{T}$. 
%
\section{Numerical methods}\label{app:numerics}
Let us write $dt$ and $du$ for the line spacings and denote the approximate value of any quantity $h(t, u)$ at the grid point $(t,u)=(t_i, u_j)$ by
\be
h^i_j \simeq h(t_i, u_j),
\ee
where $t_i \equiv t_0 + i dt$ with $i = 0, 1, \ldots$, $t_0$ is the initial time and $u_j$ is given below. We use  the following third-order Adams-Bashforth method as our time marching scheme
 \bea
 h^{i+1}_j = h^{i}_j+\f{dt}{12}\left[ 23 \dot{h}^{i}_{j} -16 \dot{h}^{i-1}_{j} + 5 \dot{h}^{i-2}_{j}\right].\label{equ:tmarch}
 \eea
In the following we discuss the choices of the grid points $u_j$ in both the Chebyshev pseudo-spectral method and the finite difference method. We will only focus on solving eq. (\ref{equ:eom}) but eq. (\ref{equ:eomEdd}) can be solved in a similar way. 
\subsection{The Chebyshev pseudo-spectral method}
In this case, the grid points $u_j$ are chosen to be
\be
u_j = (1.0 + x_j) u_{max}/2~~\text{with}~~x_j = \cos\left(\f{j \pi}{n}\right)~~\text{and}~~j=0, 1, \ldots, n. 
\ee
Accordingly, the differentiation matrix 
\be
D_{n+1} = 2 d_{n+1}/u_{max},
\ee
where the elements of the $(n+1)-$by$-(n+1)$ matrix $d_{n+1}$ are given by(\cite{withPaul} and references therein)
 \bea
         \left( d_{n+1} \right)_{ij} = \left\{
        \begin{array}{c l}
            \f{2 n^2+1}{6}& ~~~~i=j=0\\
            \f{-x_j}{2 \left(  1 -x_j^2 \right)}& ~~~~0<i = j < n\\
           \f{c_i}{c_j} \f{(-1)^{i+j}}{x_i - x_j}& ~~~~i\neq j\\
             -\f{2 n^2+1}{6}& ~~~~i=j=n\\           
        \end{array}
        \right.
\eea
with $c_0=c_n=2$ and $c_j=1$ otherwise. The difference equations are obtained by replacing $(\ldots)'$ with $D_{n+1} (\ldots)$ in (\ref{equ:eom}). Given $V$, $P$, $f$ and $\delta$ at $t_i$ and the previous time steps, all $V^{i+1}_j$ and $P^{i+1}_j$ are calculated by solving the discretized versions of (\ref{equ:Vdot}) and (\ref{equ:Pdot}) except $V^{i+1}_n$ and $P^{i+1}_0$, which are given by the boundary condition in (\ref{equ:phi0}), i.e.,
\be
V^{i+1}_n = -2 t_{i+1} \epsilon e^{-a t_{i+1}^2},~~\text{and}~~P^{i+1}_0 = 0.
\ee

\subsection{The finite difference method}
In this case, the grid points $u_j$ are chosen to be equally spaced, that is,
\be
u_j = j du + u_{min}~~\text{with}~~j= 0, 1, \ldots, n~~\text{and}~~du=\left(u_{max}-u_{min}\right)/n.
\ee
We take $u_{min} = 0.005$, which is introduced to avoid the numerically singular behavior near $u=0$\cite{Garfinkle02}. The derivatives with respect to $u$ in eq.s (\ref{equ:Vdot}) to (\ref{equ:deltap}) are replaced by finite differences
\begin{subequations}\label{equ:eomdiff}
\bea
&&\dot{V}^i_j = -\f{3}{u_j} f^i_j e^{-\delta^i_j} P^i_j +  \left( f^i_j e^{-\delta^i_j} P^i_j -  f^i_{j-1} e^{-\delta^i_{j-1}} P^i_{j-1} \right)/du,\label{equ:Vdotdiff}\\
&&\dot{P}^i_j = \left( f^i_{j+1} e^{-\delta^i_{j+1}} V^i_{j+1}  - f^i_j e^{-\delta^i_j} V^i_j \right)/du,\label{equ:Pdotdiff}\\
&&\dot{f}^i_j = \f{4}{3} u_j {f^i_j}^2 e^{-\delta^i_j} V^i_j P^i_j,\label{equ:fdotdiff}\\
&&\f{\delta^i_j - \delta^i_{j-1}}{du}=\f{2}{3} u_{j} \left(  {V^i_{j}}^2 + {P^i_{j}}^2 \right).\label{equ:deltapdiff}
\eea
\end{subequations}
With the initial conditions in (\ref{equ:initial})/(\ref{equ:adsbh}) and the boundary conditions
\be
V^{i+1}_0 = -2 t_{i+1} \epsilon e^{-a t_{i+1}^2},~~P^{i+1}_n = 0,~~\text{and}~~\delta^{i+1}_0=0,
\ee
one can calculate the late-time geometry by (\ref{equ:eomdiff}) and (\ref{equ:tmarch}). 

\end{document}